\documentclass[preprint,aps]{revtex4}
\usepackage{graphics}
\usepackage{dcolumn}
\usepackage{bm}

\begin{document}
\preprint{ }
\title{ Thermal kaon production in relativistic heavy ion collisions}
\author{I. Zakout, W. Greiner}
\affiliation{Institut f\"ur Theoretische Physik, J. W. Goethe Universita\"at, Robert-Mayer-Stra$\beta$e 8-10,
Postfach 11 19 32, D-60054 Frankfurt am main, Germany}
\author{H. R. Jaqaman}
\affiliation{Physics Department, Bethlehem University, P.O. Box 9 Bethlehem, Palestinian Authority}

\begin{abstract}

We study kaon production in hot and dense hypernuclear matter 
with a conserved zero total strangeness and a conserved small negative isospin charge fraction
in order to make our results relevant to relativistic heavy ion collisions.  
The baryons and kaons are treated as MIT bags in the 
context of the modified quark-meson coupling model.
They interact with each other via the scalar mesons 
$\sigma,\sigma^*$ and the vector mesons $\omega,\phi$ as well
as the iso-vector meson $\rho$. 
We adopt realistic sets of hyperon-hyperon interactions based on several versions
of the Nijmegen core potential models.
Our results  indicate that the hyperons as well as the kaons are produced 
abundantly when the temperature increases and approaches 
the critical temperature for the phase transition to a quark-gluon plasma. 
Moreover we find that the kaons are only produced thermally
and we find no kaon condensation in the regime explored by relativistic heavy ion collisions.
\end{abstract}


\maketitle
\section{Introduction}

The quark-meson coupling (QMC) model incorporates the quark-gluon degrees of freedom while
respecting the established model based on the hadronic degrees of freedom
in nuclei. It describes nuclear matter as a collection of non-overlapping MIT bags
interacting through the self-consistent exchange of scalar $\sigma,\sigma^*$ and vector mesons $\omega,\phi$ as well
as the iso-vector meson $\rho$ in the mean field approximation with the 
mesonic fields directly coupled to the quarks\cite{Saito94}.
In the so-called modified quark-meson coupling model 
(MQMC)\cite{Jennings96a,Jennings96b,Jennings97}, 
it has been suggested that including a medium-dependent bag parameter
is essential for the success of relativistic nuclear phenomenology.
The density-dependence of the bag parameter is introduced by coupling 
it to the scalar mesonic field\cite{Jennings96a,Jennings97}. 
We have used the MQMC to study
the properties of nuclear matter and quark deconfinement 
at finite temperature\cite{Jaqaman99a,Jaqaman99b,Jaqaman00}
and the phase transition from the hadronic phase to the quark-gluon plasma
\cite{Jaqaman00,Jaqaman01}. 
We have also studied the properties of neutron stars in the context of MQMC\cite{Pal99}.
 
In the present work we shall extend the MQMC to study the production 
of strange hadrons, especially kaons, 
in relativistic heavy ion collisions. Elucidating the properties of
kaon production and kaon condensation is important to specify the 
tricritical point of the phase transition from 
a QGP phase to a hadronic phase.
The competition between pion and kaon condensation in the framework
of a microscopic model has been considered at zero
and finite temperatures in the space of quark chemical potentials
\cite{Barducci04}. 
We consider a system of strange hadronic matter with conserved zero total strangeness
and a conserved small negative isospin charge fraction in order to make 
our results relevant to heavy ion collisions. We also investigate 
the possibility for the onset of kaon condensation in such a system.  

In the mean field approximation, 
the scalar  meson $\sigma$ with mass 550 MeV
is supposed to simulate the exchange
of correlated pairs of pions and may
represent a very broad resonance observed in $\pi\pi$ scattering, while
the vector meson $\omega$  is identified with an actual meson having a mass
of about 780 MeV. The $\sigma$ scalar and $\omega$ vector mesonic fields couple the up and down flavors.
Also the scalar $\sigma^*(s\overline{s})$   
and vector $\phi(s\overline{s})$ mesonic fields are actual mesons with masses 975 and 1020 MeV, respectively.
They are introduced to couple to the strange flavor quark.
Similarly, the mesonic iso-vector field $\rho$, which is essential for 
isospin charged nuclear matter, is an actual meson
with mass 770. 
It exchanges the non-strange flavor quarks. We assume hadronic matter to consist of the 
members of the SU(3) baryon octet and the kaon doublet. The pions are not explicitly 
included in the calculation because
the contribution of one-pion exchange vanishes in spin and isospin saturated 
nuclear matter and so their inclusion will not have any effect on the quantities of 
interest in the present work such as baryon or strangeness density. The pions contribute to the 
total energy density and to the pressure, for instance, but these quantities are not calculated 
in the present work. For the purpose of calculating these latter quantities, the pions can be 
included as an ideal gas. 
The baryon octet is comprised of 
$(p,n,\Lambda,\Sigma^+,\Sigma^0,\Sigma^-,\Xi^0,\Xi^-)$ baryons
while the kaon doublet is comprised of $(K^+,K^0)$ mesons.
In heavy ion collisions the total net strangeness of the system is 
conserved during the collision. If the quark-gluon plasma is produced during the collision
the net strangeness for the hadronic phase may not be conserved but the total net strangeness 
for both the hadronic and quark-gluon phases will be conserved. 

Various calculations and models predict that metastable strange systems with strangeness fractions of 
order one and charge neutrality might exist in
the hadronic phase at densities between two 
and three times nuclear matter saturation density\cite{Schaffner93,Schaffner94}. 
However, the predicted phenomenon of metastability of strange hadronic matter 
and the actual values of the binding energy depend specifically on the partly unknown hyperon potentials 
assumed in dense matter\cite{Schaffner93,Schaffner94,Balberg94,Wang99}.
Recently, unconstrained Relativistic Mean Field calculations of a charge-neutral strangeness-rich hadronic system
have been carried out with two sets of hyperon-hyperon (or $YY$) potentials,
the first set is determined from the Nijmegen hard-core
potential Model D\cite{Nagels77}, and the second set corresponds to the
potentials obtained in a recent SU(3) extension of the
Nijmegen soft-core potential Model NSC97\cite{Stoks99a,Stoks99b,Stoks99c,Stoks00}.
The differences between these two sets are essentially attributable to the extremely
attractive $\Sigma\Sigma$ and $\Xi\Xi$ interactions in the second set which allow for the
possibility of deeply bound nuclear matter with hyperons\cite{Gal00}.

In the present work, we shall study the production of kaons in neutral as well as weakly
isospin strange hadronic matter at finite temperature.
We shall adopt realistic phenomenological fitting parameters which are calculated 
from the potential models. 
The outline of the paper is as follows. In section II we write the equation of state. 
Then in section III we present the two different sets of fitting 
parameters for the $YY$ potentials. 
In section IV, we present and discuss our results. Finally, we summarize our conclusions
in section V.

\section{The Equation of State}

The quark field $\psi_{q}(\vec{r},t)$ inside a bag of radius $R_i$  
representing  a baryon of species $i$ satisfies the Dirac equation 
\begin{eqnarray}
\left[ i\gamma^{\mu}\partial_{\mu}- m_{q}^{0}
+(g_{q\sigma}\sigma-g_{q\omega}\omega_{\mu}\gamma^{\mu})
+(g_{q\sigma^{*}}\sigma^{*}-g_{q\phi}\phi_{\mu}\gamma^{\mu})
-g_{q\rho}\frac{1}{2}\vec{\tau}\cdot\vec{\rho}_{\mu}\gamma^{\mu}
\right]\psi_{q}(\vec{r},t)=0,
\label{Dirac}           
\end{eqnarray}
where $m_{q}^{0}$ is the current mass of a quark of flavor $q$.
The current quark masses are taken, for the up and down flavor quarks, 
to be $m_u=m_d=0$ while for the strange flavor $m_s=150\mbox{MeV}$. 
Inclusion of small current quark masses for the non-strange flavors or
other values for the strange flavor leads only to 
small numerical refinements of the present results.
In the mean field approximation 
the meson fields are treated classically and 
the space like components of the vector fields vanish for
infinite systems due to rotational invariance.
As a result we get $\omega_{\mu}\gamma^{\mu}=<\omega_0>\gamma^{0}=\omega\gamma^{0}$
and $\phi_{\mu}\gamma^{\mu}=<\phi_0>\gamma^{0}=\phi\gamma^{0}$
and 
$\vec{\tau}\cdot\vec{\rho}_{\mu}\gamma^{\mu}=
\tau_3<\rho>\gamma^{0}=\tau_3\rho\gamma^{0}=2I_{3i}\rho\gamma^{0}$.  
The non strange (up and down) flavor quarks are coupled to the scalar 
$\sigma(550)$ and vector $\omega(780)$ and iso-vector $\rho(980)$
mesons while the strange flavor quarks are coupled to 
$\sigma^{*}(975)$ and $\phi(1020)$.

For a given value of the bag radius $R_i$ and scalar fields
$\sigma$ and $\sigma^*$, the quark momentum
$x_{qi}$ is determined by the boundary condition of confinement which,
for quarks of flavor $q$ in a spherical bag, reduces to
$j_0({x_{qi}})=\beta_{qi} j_1({x_{qi}})$, where
\begin{eqnarray}
\beta_{qi}=\sqrt{ 
\frac{{\Omega_{qi}}(\sigma,\sigma^*)-R_i m^{*}_{q} }
{{\Omega_{qi}}(\sigma,\sigma^*)+R_i m^{*}_{q} }}.
\label{bleta}            
\end{eqnarray}
We have defined the effective quark mass inside the bag as 
\begin{eqnarray}
m^{*}_{q}=m^{0}_{q}-g_{q\sigma}\sigma-g_{q\sigma^{*}}\sigma^{*},
\label{effmass}         
\end{eqnarray}
and the effective quark energy is given by
\begin{eqnarray}
\Omega_{qi}(\sigma,\sigma^*)/R_i=\sqrt{ (x_{qi}/R_i)^{2}+{m_q^*}^{2} }.
\label{Omegnk}          
\end{eqnarray}
The bag energy is given by
\begin{eqnarray}
E_{\mbox{bag}_i}= E_{\mbox{quarks}_i}
- \frac{Z_i}{R_i}+\frac{4\pi}{3}R_i^{3}  B_i(\sigma,\sigma^*),
\label{Ebag}         
\end{eqnarray}
where
\begin{eqnarray}
E_{\mbox{quarks}_i}=\sum^{n_q}_q \frac{\Omega_{qi}(\sigma,\sigma^*)}{R_i}
\label{new1}
\end{eqnarray}
is the total constituent quarks kinetic energy inside the bag
and $\frac{Z_i}{R_i}$ term is the zero-point energy of the quarks
and $B_i(\sigma,\sigma^*)$ is the medium-dependent bag parameter.

We would like to emphasize here that we did not take into account 
the thermal excitation of the constituent quarks inside the 
bag\cite{Jaqaman99a,Jaqaman99b,Jaqaman00,Panda97}.
In Refs.\cite{Jaqaman99a,Jaqaman99b,Jaqaman00,Panda97},
the total constituent quarks kinetic energy with thermal excitations of the populated 
quarks is given by 
\begin{eqnarray}
E_{\mbox{quarks}_i}=\sum^{n_q}_q \frac{\Omega_{qi}(\sigma,\sigma^*)}{R_i}
\left( f_q[\epsilon^*_q-\mu_q]-f_q[\overline{\epsilon}^*_q+\mu_q] \right)
\label{Old1}
\end{eqnarray}
where $f_q[\epsilon^*_q-\mu_q]$ and $f_q[\overline{\epsilon}^*_q+\mu_q]$ 
are the Fermi-Dirac quark and antiquark distribution functions, respectively. 
The constituent quark effective energy is given by
\begin{eqnarray}
\epsilon^*_q=\frac{\Omega_{qi}(\sigma,\sigma^*)}{R_i} 
+ (g_{q\omega}\omega+g_{q\phi}\phi+\frac{1}{2}g_{q\rho}\tau_3\rho)
\end{eqnarray} 
and 
\begin{eqnarray}
\overline{\epsilon}^*_q=\frac{\Omega_{qi}(\sigma,\sigma^*)}{R_i} 
- (g_{q\omega}\omega+g_{q\phi}\phi+\frac{1}{2}g_{q\rho}\tau_3\rho)
\end{eqnarray}
In this construction, the constituent quark chemical potential
$\mu_q$ is determined from the constraint
\begin{eqnarray} 
n_q=\sum_q^{n_q} f_q[\epsilon^*_q-\mu_q]-f_q[\overline{\epsilon}^*_q+\mu_q]
\label{Old2}
\end{eqnarray} 
where $n_q$ is the number of the constituent quarks (antiquarks) inside the bag.
This constraint, fortunately, gives $f_q\approx 1$ and $\overline{f}_q\approx 0$
in the actual numerical calculations.
We have found that adopting Eq.(\ref{new1}) rather Eq.(\ref{Old1}) 
in the calculations doesn't affect the results given 
in Refs\cite{Jaqaman99a,Jaqaman99b,Jaqaman00}.
Therefore, the thermal excitations of the populated quarks inside 
the bags in that construction is negligible\cite{Jaqaman99a,Jaqaman99b,Jaqaman00}.  

In the simple QMC model, the bag parameter (constant) is taken as $B_{0}$ 
corresponding to its value for a free baryon. 
The medium effects are taken into account in the 
MQMC\cite{Jennings96a,Jennings96b} by coupling the bag parameter to the scalar 
meson fields. 
In the present work we use the following generalized ansatz for 
the coupling of the bag parameter to the scalar fields 
$\sigma$ and $\sigma^*$ \cite{Jaqaman00,Pal99,Jaqaman01}:
\begin{eqnarray}
B_i(\sigma,\sigma^*)=B_{0}\exp
\left[-4
\left(g^{\mbox{bag}}_{i\sigma}\sigma
+g^{\mbox{bag}}_{i\sigma^*}\sigma^*)/M_{i}\right)
\right]
\label{bagCon}          
\end{eqnarray}
with $g^{\mbox{bag}}_{i\sigma}$ and $g^{\mbox{bag}}_{i\sigma^*}$ 
as additional coupling constants to be fitted from the phenomenology. 
In Ref.\cite{Wang99}, 
the bag parameter has been considered to couple to the non strange $\sigma$ scalar field 
but not to the strange $\sigma^{*}$ scalar field.  
The spurious center-of-mass energy is subtracted to obtain the effective baryon mass
\begin{eqnarray}
M^{*}_{i}=\sqrt{{E_{\mbox{bag}_i}}^2 - <p^{2}_{\mbox{cm}}>_i},
\label{MNSTAR}          
\end{eqnarray}
where 
\begin{eqnarray}
<p^{2}_{\mbox{cm}}>_i=
\sum^{n_q}_q x_{qi}^2 / R_i^2.
\label{PCM}            
\end{eqnarray}
The bag radii $R_{i}$ and $R_k$ for baryon $i$ and kaon $k$ species, respectively,
are obtained through the minimization of the baryon mass with respect to the bag 
radius\cite{Saito94}
\begin{eqnarray}
\frac{\partial M^{*}_{i}}{\partial R_{i}}=0.
\label{MNR}     
\end{eqnarray}

The coupling of the scalar mean fields $\sigma$ and $\sigma^*$ with
quarks in the non overlapping MIT bag through the solution
of the point-like Dirac equation should be taken into account
to satisfy the self-consistency condition.  
However, this constraint is essential to obtain the correct solution of the 
scalar mean field $\sigma$ and $\sigma^*$. We think the discrepancy between
our results in Ref.\cite{Jaqaman99a} and those in Ref.\cite{Panda97} is due
to the coupling of the quark with 
the scalar mean fields $\sigma, \sigma^*$ in the framework of the point 
like Dirac equation.
The differentiation of the effective hadron species mass with respect to
$\sigma$ gives
\begin{eqnarray}
\frac{\partial {M^*_i}}{\partial \sigma}=
\frac{
E_{\mbox{bag}_i}\left(\frac{\partial E_{\mbox{bag}_i}}{\partial \sigma}\right)_{\sigma^*}
-\frac{1}{2}
\left(\frac{\partial <p^2_{\mbox{cm}}>_i}{\partial \sigma}\right)_{\sigma_*} 
}
{M^*_i}
\label{Dirac_point}
\end{eqnarray}
where
\begin{eqnarray}
\left(
\frac{\partial E_{\mbox{bag}_i}}{\partial \sigma}
\right)_{\sigma^*}=
\sum_q 
\left(
\frac{\partial E_{\mbox{bag}_i}}
{\partial \Omega_{qi}}
\right)
\left(
\frac{\partial \Omega_{qi}}
{\partial \sigma}
\right)
+
\left(
\frac{\partial E_{\mbox{bag}_i}}{\partial \sigma}
\right)_{\Omega_{qi}}
\end{eqnarray}
\begin{eqnarray}
\frac{\partial E_{\mbox{bag}_i}}{\partial \Omega_{qi}}=\sum_{q'}
\frac{\delta_{q'q}}{R_i}
\end{eqnarray}
and           
\begin{eqnarray}
\left(
\frac{\partial E_{\mbox{bag}_i}}{\partial \sigma}
\right)_{\Omega_{qi}}=
\frac{4\pi}{3}R^3_i \frac{\partial B_i(\sigma,\sigma^*)}{\partial \sigma}
\end{eqnarray}
The $\frac{\partial \Omega_{qi}}{\partial \sigma}$ depends on
$x_{qi}$. 
Its solution is obtained from the point-like Dirac equation for the quarks 
and must satisfy the required 
boundary condition as well as the confinement on the bag surface\cite{Saito94,Jennings96a}. 
It reads
\begin{eqnarray}
\frac{1}{R_i}\frac{\partial \Omega_{qi}}{\partial \sigma}=
-g_{q \sigma}<\overline{\psi}_i|\psi_i>,
\end{eqnarray}
where
\begin{eqnarray}
<\overline{\psi}_i|\psi_i>=\frac{\Omega_{qi}/2+m^*_q R_i(\Omega_{qi}-1)}
{\Omega_{qi}(\Omega_{qi}-1)+m^*_q R_i/2}.
\end{eqnarray}
Furthermore, the  term for the spurious center-of-mass momentum 
correction reads
\begin{eqnarray}
\frac{\partial <p^2_{cm}>_i}{\partial \sigma}=
\frac{1}{R^2_i}\sum_q \frac{\partial x^2_{qi}}{\partial \sigma},
\end{eqnarray}
where
\begin{eqnarray}
\frac{\partial x^2_{qi}}{\partial \sigma}
=\left(2\Omega_{qi}
\frac{\partial\Omega_{qi}}{\partial\sigma}\right)
+2 R^2_i m^*_q g_{qi}.
\end{eqnarray}

Similar expressions can be obtained for the differentiation of the effective 
hadron (baryon and kaon) species mass with respect to the scalar mean field $\sigma^*$.
The zero-point energy parameters  $Z_i$ of Eq.(4) are used to fit the 
actual masses of the free baryons 
$M_i=939, 1116, 1193$ and $1315 \mbox{MeV}$
and are found to be 
$Z_i$=2.03, 1.814, 1.629 and 1.505 for the $N, \Lambda, \Sigma$ 
and $\Xi$ hyperons respectively, corresponding to a 
free baryon bag parameter $B_0=(188.1)^4 \mbox{MeV}^{4}$
and a free nucleon bag radius $R_0=0.6 \mbox{fm}$. 
The zero-point energy parameter $Z_k$ is determined by fitting the actual mass of the free
kaon $M_k=495$ MeV which yields the value $Z_k$=1.170.


The effective Fermi-energy for baryon and antibaryon species $i$ is given by 
\begin{eqnarray}
\epsilon^{*}_i(p)=\sqrt{ {p}^{2}+{M^{*}_{i}}^{2} }+X_i
\end{eqnarray}
and
\begin{eqnarray}
\overline{\epsilon}^{*}_i(p)=\sqrt{ {p}^{2}+{M^{*}_{i}}^{2} }-X_i,
\end{eqnarray}
respectively, where
\begin{eqnarray}
X_i=[g_{i\omega}\omega + g_{i\phi}\phi + g_{i\rho}{I_3}_{i}\rho].
\end{eqnarray}
The effective Bose-energy for the kaon and antikaon species reads
\begin{eqnarray}
\epsilon^{*}_k(p)=\sqrt{ {p}^{2}+{M^{*}_{k}}^{2} }+X_k,
\end{eqnarray}
and
\begin{eqnarray}
\overline{\epsilon}^{*}_k(p)=\sqrt{ {p}^{2}+{M^{*}_{k}}^{2} }-X_k,
\end{eqnarray}
respectively, where
\begin{eqnarray}
X_k=[g_{k\omega}\omega + g_{k\phi}\phi + g_{k\rho}{I_3}_{k}\rho].
\end{eqnarray}
The indices for the kaon and antikaon species are given by
$k=(K^+,K^0)$ and $\overline{k}=(K^-,\overline{K}^0)$, respectively.

The chemical potentials $\mu_i$ and $\mu_k$ for baryon species $i$ 
and kaon species $k$ read
\begin{eqnarray}
\mu_i=B_i\mu_B+S_i\mu_S+Q_i\mu_Q,
\label{b_chemical}
\end{eqnarray}
and
\begin{eqnarray}
\mu_k=S_k\mu_S+Q_k\mu_Q,
\label{k_chemical}
\end{eqnarray}
respectively.
The quantum numbers 
$B_i$, $S_i$ and $Q_i$ are the baryon, strangeness and isospin quantum numbers, respectively.
The baryon, strangeness, isospin chemical potentials $\mu_B$,
$\mu_S$ and $\mu_Q$, respectively, are determined from the constraints of
total baryon, strange and isospin densities $n_B$ Eq.(\ref{n_baryon}),
$n^{\mbox{Total}}_{S}$ Eq.(\ref{n_strange}) 
and $n^{\mbox{Total}}_{Q}$ Eq.(\ref{n_isospin}), respectively.

The density for each baryon and antibaryon species is given by
\begin{eqnarray}
n_i=\frac{d_i}{(2\pi)^{3}}
\int d^{3}p f[\epsilon^{*}_i(p)-\mu_i],
\label{b_dens}
\end{eqnarray}
and
\begin{eqnarray}
\overline{n}_i=\frac{d_i}{(2\pi)^{3}}
\int d^{3}p f[\overline{\epsilon^{*}_i}(p)+\mu_i],
\label{antib_dens}
\end{eqnarray}
respectively, where $d_i=2$ is the spin degeneracy for the octet baryon 
species and $f$ is 
the Fermi-Dirac distribution function 
\begin{eqnarray}
f[x]=\frac{1}{(e^{x/T}+1)}.
\end{eqnarray}
Similarly, the density for each kaon and antikaon species is given by
\begin{eqnarray}
n_k=\frac{d_k}{(2\pi)^{3}}
\int d^{3}p b[\epsilon^{*}_k(p)-\mu_k],
\label{k_dens}
\end{eqnarray}
and
\begin{eqnarray}
\overline{n}_k=\frac{d_k}{(2\pi)^{3}} 
\int d^{3}p b[\overline{\epsilon^{*}_k}(p)+\mu_k],
\label{antik_dens}
\end{eqnarray}
respectively, where $d_k=1$ is the kaon spin degeneracy, and  $b$ 
is the Bose-Einstein distribution function
\begin{eqnarray}
b[x]=\frac{1}{(e^{x/T}-1)}.
\end{eqnarray}

When the condition for kaon condensation (KC)\cite{Kaplan86,Nelson87} is not satisfied, 
see Eq.(\ref{c_condensate}) below, the kaons will not condensate 
and the density of the kaon condensate $n_{\mbox{KC}}=0$. 
In this case the chemical potentials $\mu_B,\mu_S$ and $\mu_Q$ are determined
from $n_B, n^{\mbox{Total}}_S$ and $n^{\mbox{Total}}_Q$. 
However, when the condition for the kaon condensation 
$\mu_K=\epsilon^{*}_K(0)$ is satisfied for either 
kaon species $K^+$ or $K^0$ then that species will condensate.
Beyond the condensation point, the strange chemical potential saturates 
at the threshold energy for the ground states $\mu_K\equiv\epsilon^{*}_K(0)$.
On the other hand if the condition $\mu_K=-\overline{\epsilon}^{*}_K(0)$ is satisfied 
for either $K^-$ or $\overline{K}^0$ antikaon, then either $K^-$ or $\overline{K}^0$ 
antikaon will condensate. Beyond the condensation point, 
the strange chemical potential saturates 
at the threshold energy for the ground states $\mu_K\equiv-\overline{\epsilon}^{*}_K(0)$.
We shall return to the kaon condensation when we reach Eq.(\ref{c_condensate}).

The baryon density reads,
\begin{eqnarray}
n_{B}=\sum^{\mbox{Bs}}_i B_i [n_i-\overline{n}_i].
\label{n_baryon}
\end{eqnarray}
The total strangeness density is given by
\begin{eqnarray}
n_{S}=n^{\mbox{Bs}}_{S}+n^{\mbox{Ks}}_{S},
\label{n_strange}
\end{eqnarray}
where $n^{\mbox{Bs}}_{S}=\sum^{\mbox{Bs}}_i S_i [n_i-\overline{n}_i]$
and
$n^{\mbox{Ks}}_{S}=\sum^{\mbox{Ks}}_k S_k [n_k-\overline{n}_k]+S_K  n_{\mbox{KC}}$.
The total isospin density is given by
\begin{eqnarray}
n_{Q}=n^{\mbox{Bs}}_{Q}+n^{\mbox{Ks}}_{Q},
\label{n_isospin}
\end{eqnarray}
where $n^{\mbox{Bs}}_{Q}=\sum^{\mbox{Bs}}_i Q_i [n_i-\overline{n}_i]$ 
and 
$n^{\mbox{Ks}}_{Q}=\sum^{\mbox{Ks}}_k Q_k [n_k-\overline{n}_k]+Q_K  n_{\mbox{KC}}$.

The vector mean fields $\omega$, $\phi$ and $\rho$
are determined, respectively, as follows
\begin{eqnarray}
\omega&=&
\sum^{\mbox{Bs}}_i \frac{g_{i\omega}}{m^{2}_{\omega}} 
[n_{i}-\overline{n}_i]
+
\sum^{\mbox{Ks}}_k \frac{g_{k\omega}}{m^{2}_{\omega}}
[n_{k}-\overline{n}_k]
+
\frac{g_{K\omega}}{m^{2}_{\omega}} n_{\mbox{KC}},
\label{v_omega}
\\
\phi&=&
\sum^{\mbox{Bs}}_i \frac{g_{i\phi}}{m^{2}_{\phi}}
[n_{i}-\overline{n}_i]
+
\sum^{\mbox{Ks}}_k \frac{g_{k\phi}}{m^{2}_{\omega}}
[n_{k}-\overline{n}_k]
+
\frac{g_{K\phi}}{m^{2}_{\phi}}n_{\mbox{KC}},
\label{v_phi}
\\
\rho&=&
\sum^{\mbox{Bs}}_i \frac{g_{i\rho}}{m^{2}_{\rho}}
{I_3}_i[n_{i}-\overline{n}_i]
+
\sum^{\mbox{Ks}}_k \frac{g_{k\rho}}{m^{2}_{\rho}}
{I_3}_k[n_{k}-\overline{n}_k]
+
\frac{g_{K\rho}}{m^{2}_{\rho}}
{I_3}_k n_{\mbox{KC}}.
\label{v_rho}
\end{eqnarray}
The $g_{i\omega}$, $g_{i\phi}$ and $g_{i\rho}$
are the meson-baryon coupling constants while
$g_{k\omega}$, $g_{k\phi}$ and $g_{k\rho}$
are the meson-kaon coupling constants and are determined from the data.

The pressure for hot and dense strange nuclear matter 
is equal to the negative of the grand thermodynamic potential 
density and is given by
\begin{eqnarray}
P&=&
\frac{1}{3}\sum^{\mbox{Bs}}_i\frac{d_i}{(2\pi)^{3}}\int d^{3} p
\frac{p^{2}}{\epsilon_i^{*}}
\left(f[\epsilon^*_i(p)-\mu_i]
+f[\overline{\epsilon^*_i}(p)+\mu_i]\right)
+P_{K}
\nonumber\\
&~&
+\frac{1}{2}m^{2}_{\omega}\omega^{2}
+\frac{1}{2}m^{2}_{\phi}\phi^{2}
+\frac{1}{2}m^{2}_{\rho}\rho^{2}
-\frac{1}{2}m^{2}_{\sigma}\sigma^{2}
-\frac{1}{2}m^{2}_{\sigma^*}{\sigma^*}^{2},
\label{Pres1}
\end{eqnarray}
where summation $i$ runs over the 8 species of the baryon octet
$p,n$, $\Lambda$, $\Sigma^+,\Sigma^0,\Sigma^-$ 
and $\Xi^{0},\Xi^{-}$.
The thermal  kaon pressure is given by 
\begin{eqnarray}
P_{K}=\frac{1}{3}
\sum^{\mbox{Ks}}_k \frac{d_k}{(2\pi)^{3}}
\int d^{3}k \frac{k^2}{\sqrt{k^{2}+{M^*_k}^{2}}}
\left(b[\epsilon^*_k(k)-\mu_k]
+b[\overline{\epsilon^*_k}(k)+\mu_k]\right),
\end{eqnarray}
where the summation $k$ run over the kaon doublet species $K^+,K^0$.

The scalar mean fields $\sigma$ and $\sigma^*$ are calculated by maximizing 
the pressure 
\begin{eqnarray}
\frac{\partial P}{\partial \sigma}&=&0,
\label{dP1}
\\
\frac{\partial P}{\partial \sigma^*}&=&0.
\label{dP2}
\end{eqnarray}
The pressure depends explicitly on the scalar mean fields $\sigma,\sigma^*$ through the
last two terms in Eq.(\ref{Pres1}). It also depends 
on the baryon and kaon effective masses $M^*_i$ and $M^*_k$ which in turn
depend on $\sigma$ and $\sigma^*$. If we write the pressure as a function of 
$(\{M^*_i\},\{M^*_k\})$ and $(\sigma,\sigma^*)$, the extremization of 
$P(\{M^*_i\},\{M^*_k\},\sigma,\sigma^*)$ with respect to the scalar fields $\sigma$
and $\sigma^*$ can be written as
\begin{eqnarray}
\frac{\partial P}{\partial \sigma}=
\sum_i \left( \frac{\partial P}{\partial M^*_i}\right)_{{\{M_j\}}_{j\ne i},\mu_B,\mu_S\mu_Q,T}
\frac{\partial M^*_i}{\partial \sigma} + 
\left(\frac{\partial P}{\partial \sigma}\right)_{\{M_j\}}=0.
\label{dPself1}
\end{eqnarray}
A similar expression can be written for $\frac{\partial P}{\partial \sigma^*}=0$.
The coupling of the scalar mean fields $\sigma,\sigma^*$ with the quarks in  
the non-overlapping MIT bags through the solution of the point-like Dirac equation
should be taken into account to satisfy the self-consistency condition. 
This constraint is essential to obtain the correct solution of 
the scalar mean field $\sigma,\sigma^*$. 

The pressure of the kaon condensate $P_{\mbox{KC}}$, if it exists, contributes 
to the total pressure of hot and dense strange hadronic matter.
The total pressure of the system becomes
\begin{eqnarray}
P^{\mbox{total}}=P+P_{\mbox{KC}}.
\end{eqnarray}
The contribution of the kaon condensate pressure reads
\cite{Glendenning98,Glendenning99}
\begin{eqnarray}
P_{\mbox{KC}}=
\frac{1}{2}\left(f\theta\right)^2
[\epsilon^{*}_k(0)-\mu_k][\overline{\epsilon^{*}_k}(0)+\mu_k],
\label{P_KC}
\end{eqnarray}
where
\begin{eqnarray}
\epsilon^{*}_k(0)=M^{*}_{k}+X_k
\end{eqnarray}
\begin{eqnarray}
\overline{\epsilon^{*}_k}(0)=M^{*}_{k}-X_k.
\end{eqnarray}
This equation is derived by assuming that the kaon amplitude 
is replaced by the ansatz $K=f\theta/\sqrt{2}$ and $K\overline{K}=\frac{1}{2}f^2\theta^2$
where $f$ is the kaon decay constant and $\theta$ is the dimensionless kaon field strength.
The variation of Equation (\ref{P_KC}) gives
\begin{eqnarray}
\delta P_{\mbox{KC}}=
f^2\theta
[\epsilon^{*}_k(0)-\mu_k][\overline{\epsilon^{*}_k}(0)+\mu_k]
=0.
\label{c_condensate}
\end{eqnarray}
There are three solutions to Eq.(\ref{c_condensate}) that determine the conditions for kaon condensation.
The trivial solution is that kaon amplitude vanishes $\theta=0$.
This case means that no kaon condensation takes place in the system and subsequently $n_{\mbox{KC}}=0$.
The other two solutions are determined as follows:
if $[\epsilon^{*}_k(0)-\mu_k]=0$ either $k$=($K^+$ or $K^0$)
will condensate, and
if $[\overline{\epsilon^{*}_k}(0)+\mu_k]=0$
either antikaon $k=(K^- \mbox{or} \overline{K}^0)$ will condensate.
The density of the kaon condensate is given by,
\begin{eqnarray}
n_{\mbox{KC}}=\left(f\theta\right)^2\left(\mu_k-X_k\right).
\label{c_amplitude}
\end{eqnarray}
In this case, the charge chemical potential becomes 
$\mu_Q(\epsilon^*_K(0),\mu_S)$ for either $K^+$or $K^0$ condensate
or 
$\mu_Q(-\overline{\epsilon}^*_K(0),\mu_S)$ for either $K^-$ or
$\overline{K}^0$ condensate since $\mu_K=S_K\mu_S+Q_K\mu_Q$ and $n^{\mbox{Total}}_S$ is conserved. 
It is interesting to note here that the three solutions give 
$P_{\mbox{KC}}=0$ and normally the pressure of the condensate kaons contributes
indirectly via the strange hadronic abundances and the total energy density of the system.

\section{Fitting parameters for $YY$ potentials}


We assume that the  $\sigma$ and $\omega$ mesons 
couple only to the up and down quarks
while $\sigma^*$ and $\phi$ couple to the strange quark.
We thus set $g_{r\phi}=g_{r\sigma^{*}}=g_{s\sigma}= g_{s\omega}=0$ 
and $g_{s\rho}=0$ where $r$ refers to the up
and down flavors while $s$ denotes the strange flavor.
By assuming the $SU(6)$ symmetry of the simple quark model
we have the relations $g_{s\sigma^*}=\sqrt{2}g_{r\sigma}$ and 
$g_{s\phi}=-\sqrt{2}g_{r\omega}$.
The $\sigma$ mean field is supposed to simulate
the exchange of correlated pairs of pions and may represent
a very broad resonance observed in $\pi \pi$
scattering. We take  $m_\sigma=550$ MeV.
The vector $\omega$ and $\rho$ mesons are identified with
the actual mesons whose masses are $m_\omega=783$ MeV
and $m_\rho=770$ MeV, respectively.
Since the mean fields, $\sigma$ and $\omega$, are considered as
$<u\overline{d}>$ condensates, they interact only with
$u,d$-quark in the baryons and kaons.
On the other hand, the scalar and vector mean fields
$\sigma^*, \phi$ are considered  as $<s\overline{s}>$ condensates
and interact only with the $s$-quarks in the baryons and kaons.
The iso-vector mean fields interact with the $u,d$-quarks in the baryons and 
kaons.

The coupling of each baryon species with the vector mesons 
is calculated by counting the constituent quarks
\begin{eqnarray}
g_{i\omega}=\sum_q^{3} g_{q\omega}=\sum_r g_{r\omega},
\end{eqnarray}
and 
\begin{eqnarray}
g_{i\phi}=\sum_q^3 g_{q\phi}=\sum_s g_{s\phi},
\end{eqnarray}
while
\begin{eqnarray}
g_{i\rho}=g_{q\rho}.
\end{eqnarray}
With these assumptions the only free parameters left at our disposal are
the quark-meson coupling constants $g_{r\sigma}$ and $g_{r\omega}$
and the bag coupling constants
$g^{\mbox{bag}}_{i\sigma}, g^{\mbox{bag}}_{i\sigma^{*}}$ for the 4 
baryon species and these parameters 
are adjusted to fit  nuclear properties
as well as the extrapolated properties of hypernuclear matter.
The coupling constants of the scalar and vector mesons
to the nonstrange quarks
are taken as $g_{r\sigma}=1$ and $g_{r\omega}=2.705$ 
which together with a bag coupling constant
$g^{\mbox{bag}}_{N\sigma}=6.81$  yield a
binding energy of 16 MeV and a compressibility $K^{-1}_V$ of
289 MeV at the normal saturation
density $\rho_0=0.17 \mbox{fm}^{-3}$ of nuclear 
matter\cite{Jennings96a,Jaqaman99a,Jaqaman99b}. 
The isospin vector meson coupling constant for the $\rho$-meson is taken
as $g_{q\rho}=8.086$ to reproduce the bulk symmetry energy.
The coupling constants $g^{\mbox{bag}}_{i\sigma}$ 
and $g^{\mbox{bag}}_{i\sigma^*}$ where $i=\Lambda,\Sigma,\Xi$
are determined from the $YY$ potentials assumptions.

In the present calculation, we have considered two
different models for the hyperon-hyperon interactions. 
Table I summarizes the values of the two sets of
coupling constants corresponding to 
these two models. 
The basic quark-meson coupling constants as well as the 
parameters $g^{\mbox{bag}}_{i\sigma}$ are identical 
in the two models and are chosen to fit nuclear and hypernuclear 
properties. In particular, the parameters 
$g^{\mbox{bag}}_{i\sigma}$ are determined by fitting
the hyperon potentials in nuclear matter\cite{Schaffner94}:
\begin{eqnarray}
U^{(N)}_\Lambda (\rho_0)=-30 MeV, \nonumber \\
U^{(N)}_\Sigma (\rho_0)=+30 MeV, \nonumber \\
U^{(N)}_\Xi (\rho_0)=-18 MeV, \nonumber \\
\end{eqnarray}
where the hyperon potentials are defined by
\begin{eqnarray}
U^{(i)}_i=(M^{*}_i-M_i)+(g_{i\omega}\omega+g_{i\phi}\phi).
\end{eqnarray}
However, the two models differ in their choices for the
constants $g^{\mbox{bag}}_{i\sigma^{*}}$.

Model (I) is designed to mimic the consequences of the Nijmegen hard-core
potential model D \cite{Nagels77,Schaffner94,Balberg94}. 
It is nevertheless
constrained by $\Lambda$ and $\Xi$ nuclear phenomenology, and by a few
$\Lambda\Lambda$ hypernuclei reported to date. It accounts more
realistically for the attractive $\Lambda\Lambda$ and $N\Xi$ interactions,
but ignores the $\Sigma$ hyperons altogether.
In this model,
the medium constants $g^{\mbox{bag}}_{i\sigma^{*}}$ are 
adjusted so that the potential of a single hyperon 
embedded in a bath of $\Xi$ matter becomes
\begin{eqnarray}
U^{\Xi}_{\Xi}(\rho_0)=U^{\Xi}_{\Lambda}(\rho_0)=-40 MeV  
\end{eqnarray}
in accordance with the attractive hyperon-hyperon interaction
of the Nijmegen hard core potential Model D\cite{Schaffner94,Wang99}.
Furthermore, we adopt the approximation
$U^{\Xi}_{\Xi}(\rho_0)\approx U^{\Xi}_{\Sigma}(\rho_0)$ 
to fit the medium constants.
The resulting $U^{(\Lambda)}_{\Lambda}(\rho_0/2)$
is about -20 MeV.

Model (II) is designed to generate qualitatively similar
baryon potentials to those obtained in the BHF approximation from the
SU(3) extensions of the Nijmegen soft-core potential 
model NSC97\cite{Stoks99a,Stoks99b,Stoks99c,Stoks00}. The phenomenology in 
this model departs substantially from that in Model I. The
NSC97 model has been tuned up to reproduce certain characteristics of
$\Lambda$ hypernuclei, particularly its version NSC97f. It yields
particularly attractive $\Xi\Xi$, $\Sigma\Sigma$ and $\Sigma\Xi$
interactions. Some of the shortcomings of this model are the vanishingly
weak $\Lambda\Lambda$ and $N\Xi$ interactions which are in contradiction
with the little experimental evidence available from $\Lambda\Lambda$
hypernuclei and from $\Xi$-nucleus interaction.
In this model we adjust the bag constants 
$g^{\mbox{bag}}_{i\sigma^*}$  to
reproduce qualitatively the binding energy
curve of each hyperon species in its own hyperonic matter $B^i_i$
as those produced by the Model NSC97f\cite{Stoks99c,Gal00}.
No binding occurs for $\Lambda$ hyperons
while $\Sigma$ matter is deeply bound at -33 MeV per baryon 
at ${\rho_\Sigma}_0$ which is twice as deep as ordinary nuclear matter, 
and $\Xi$ matter has an energy of -23 MeV per baryon at ${\rho_\Xi}_0$.


The kaon coupling constants are determined from
the kaon potentials when embedded in nuclear matter
$U^{(N)}_{K^+}=20$ MeV \cite{Tsushima97} 
and $U^{(N)}_{K^-}=-120$ to $-180$ MeV\cite{Glendenning99,Glendenning98,Thorsson94}.
In Model I, we have used two slightly different versions of these coupling constants. 
In set (Ia) we have taken $g_{K\omega}=g_{q\omega}$ based 
on simple $SU(6)$ symmetry while in set (Ib) we have taken 
$g_{K\omega}=1.7 g_{q\omega}$ in order to fit the kaon potential depths
\cite{Tsushima97}. In both sets we take 
 $g^{\mbox{bag}}_{K\sigma}=2.27$ and
$g^{\mbox{bag}}_{K\sigma^*}=\sqrt{2}g^{\mbox{bag}}_{K\sigma}$.
In Model II, we only take $g_{K\omega}=g_{q\omega}$ based on simple $SU(6)$ symmetry.
In our calculations we shall adopt $g_{K\omega}= g_{q\omega}$
as the normal case and we shall consider the case 
$g_{K\omega}=1.7 g_{q\omega}$ as a special case. Here it may be worth mentioning that  
recent results \cite{Tolos03} indicate that 
$U^{(N)}_{K^-} = -50$ to $-60 $ MeV and for such a shallow kaon 
potential the renormalization factor need not be as large as $1.7$.

\begin{table}
\caption{Fitting parameters}
\label{states}
\begin{tabular}[b]{cccccccccccccc}
Fit set & $g_{q\sigma}$ & $g_{q\omega}$ & $g_{q\rho}$
& $g^{\mbox{bag}}_{N\sigma}$ &
$g^{\mbox{bag}}_{\Lambda\sigma}$ & $g^{\mbox{bag}}_{\Lambda\sigma^{*}}$ &
$g^{\mbox{bag}}_{\Sigma\sigma}$ & $g^{\mbox{bag}}_{\Sigma\sigma^{*}}$ &
$g^{\mbox{bag}}_{\Xi\sigma}$ & $g^{\mbox{bag}}_{\Xi\sigma^{*}}$ &
$g^{\mbox{bag}}_{K\sigma}$ & $g^{\mbox{bag}}_{K\sigma^*}$ & $g_{K\omega}$\\
\hline
Model(Ia)
&1.0& 2.705 & 8.086 & 6.81 & 4.22 & 5.45 & 1.63 & 7.26 & 2.27 & 9.12 &
2.27 & $\sqrt{2}\times 2.27$ & $g_{q\omega}$\\
Model(Ib)
&1.0& 2.705 & 8.086 & 6.81 & 4.22 & 5.45 & 1.63 & 7.26 & 2.27 & 9.12 &
2.27 & $\sqrt{2}\times 2.27$ & $1.7g_{q\omega}$\\
Model(II)
&1.0& 2.705 & 8.086 & 6.81 & 4.22 & 0.0 & 1.63 & 10.28 & 2.27 & 10.17&
2.27 & $\sqrt{2}\times 2.27$ & $g_{q\omega}$\\
\end{tabular}
\end{table}

\section{Results and Discussions}
 
We have studied hadronic matter with a conserved strangeness density
$n_S=0$ at finite temperature using the MQMC model
which takes into account the quark degrees of freedom as well as
the medium dependence of the bag constant $B$.
We have chosen a direct coupling of the bag parameter $B$ to
the scalar mean fields $\sigma$ and $\sigma^*$ and the bag parameter
becomes $B\equiv B(\sigma,\sigma^*)$. The quarks and anti-quarks confined in the bags 
are coupled to the scalar mean fields $\sigma$ and $\sigma^*$ and the vector mean fields 
$\omega$ and $\phi$ and iso-vector mean field $\rho$.
At first, we have studied the symmetric hot and dense strange hadronic system 
of baryons and kaons with a neutral isospin charge $n_Q/n_B=0$ and 
with a conserved zero net strangeness $n_S=0$. 
Then, we have considered the asymmetric system
with a conserved isospin charge $n_Q/n_B$ and a conserved zero net strangeness.
This is relevant for heavy ion collisions where it is
possible to have an asymmetric system with a conserved low isospin 
fraction in the range $n_Q/n_B=0.0$ to $-0.2$.

The starting point of the calculation is the determination of
the chemical potentials $\mu_B, \mu_S$ and $\mu_Q$ 
using the baryon, strange and isospin densities given 
by Eqs. (\ref{n_baryon}), (\ref{n_strange}) and (\ref{n_isospin}), respectively. 
The vector mean-fields $\omega$ and $\phi$ and the iso-vector mean field $\rho$ 
are computed from Eqs.(\ref{v_omega}), (\ref{v_phi}) and (\ref{v_rho}), respectively.
Once the values for the chemical potentials $(\mu_B,\mu_S,\mu_Q)$ are given,
the values for the scalar mean fields $\sigma$ and $\sigma^*$ 
are calculated by maximizing the pressure using Eqs.(\ref{dP1}) and (\ref{dP2}), 
respectively.
These constraints should take into account the coupling of the quark with
the scalar mean fields in the frame of the point-like Dirac equation
exactly\cite{Saito94,Jennings96a,Jennings96b} to obtain consistent results.
The bag radius for each hadron species 
is obtained by minimizing the hadron mass with respect to its 
bag radius, Eq. (\ref{MNR}).

\begin{figure}
\includegraphics{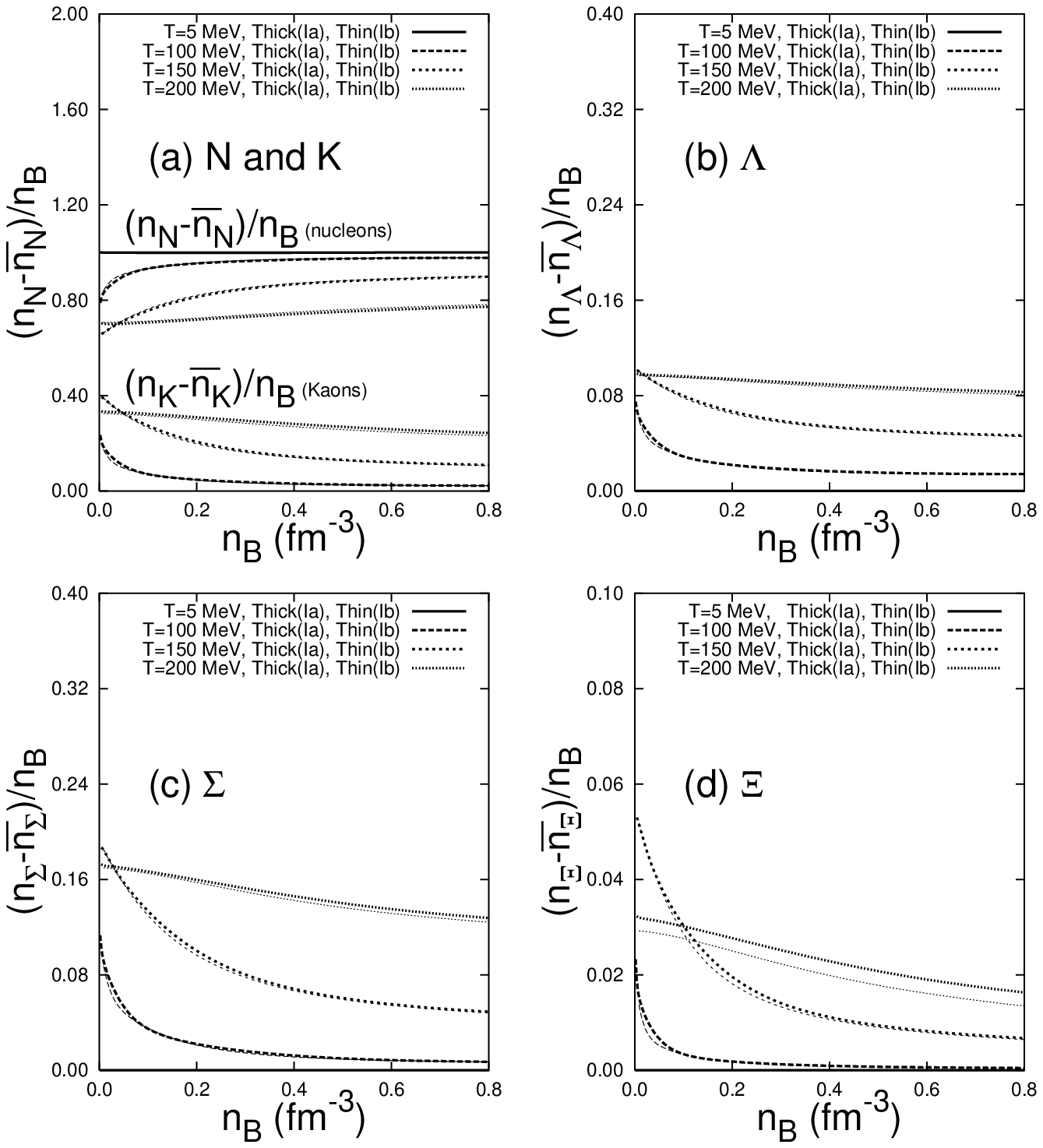}%
\caption{\label{fign1}
The density fraction for hadron and anti-hadron species 
$x_i=(n_i-\overline{n}_i)/n_B$
versus baryonic density $n_B$ at several temperatures using
the parameter sets (Ia) and (Ib).
a) Nucleons and Kaons,
b) $\Lambda$,
c) $\Sigma$,
d) $\Xi$.}
\end{figure}

In Fig.(\ref{fign1}), we display the density fraction for each baryon species
$x_i=(n_i-\overline{n}_i)/n_B$ versus the baryonic density $n_B$ at several temperatures
using the parameter sets (Ia) and (Ib). Furthermore we also 
display in the lower part of panel (a) of this 
figure the relative abundance of the kaons. The abundance of the strange 
hadrons up to $T=50 MeV$ is completely negligible and this explains why the 5 MeV curve cannot 
be distinguished from the horizontal axes of the panels (b), (c) and (d). 
The strange hadrons make their first noticeable appearance 
at $T\approx 80$ MeV and 
they start to be produced abundantly when the temperature exceeds $100 \mbox{MeV}$.
The $\Sigma$ appears to be produced more abundantly than the $\Lambda$ 
only because it has a higher isospin degeneracy factor.
On the other hand, the $\Xi$ has the lowest rate of production.
The dependence of the production rates of the various hadronic species on the total baryon 
density  seems to undergo a drastic change at the critical temperature 
$T_c=170 \mbox{MeV}$ of the QGP phase transition, becoming almost flat
above this temperature. We also note that the modification of the kaon vector 
coupling which leads to the two sets (Ia) and (Ib) has very little effect 
on the production of the various hadronic species.
The effect can be mainly seen above $T_c$ where set (Ia) produces slightly more $\Xi$ 
and, to a lesser extent, more $\Sigma$ and kaons than set (Ib) because of
the stronger vector coupling for kaons in (Ib).
Kaon  production in the system increases with temperature 
especially for temperatures exceeding $T\approx 100$ MeV. 
We see that a large mesonic strangeness  $n^{K}_S>0.2$ can be reached 
at the critical temperature $T_c = 170 \mbox{MeV}$ when kaons 
are produced abundantly in the system.

\begin{figure}
\includegraphics{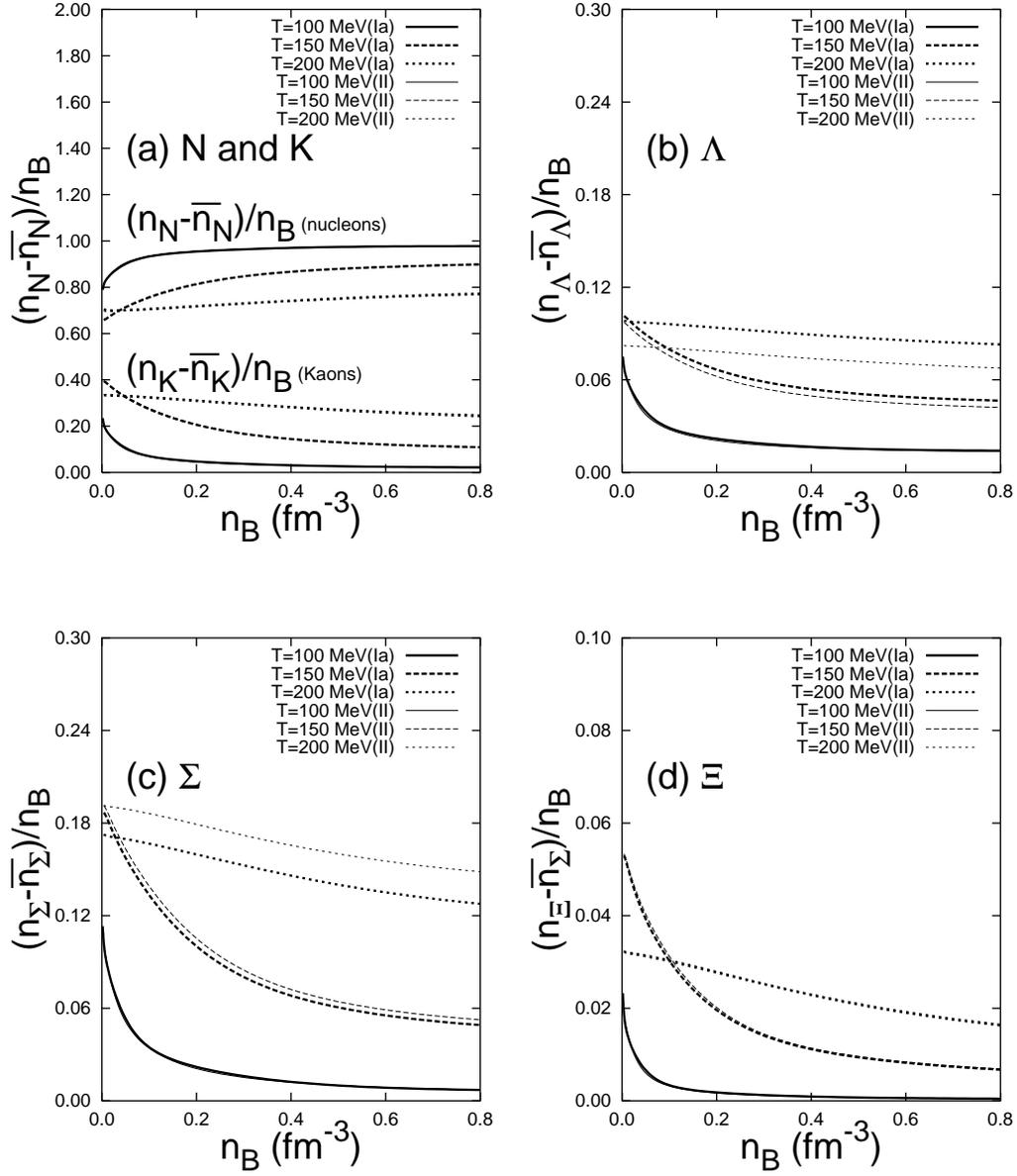}%
\caption{\label{fign2}
Same as Fig.(\ref{fign1}) but using parameter sets (Ia) and (II).}
\end{figure}

In Fig.(\ref{fign2}), we compare the baryon and kaon abundances 
obtained using parameter sets (Ia) and (II).  
The phenomenology in model (II) departs substantially from that 
in (Ia) for large baryonic and strangeness densities
where the phase transition to $\Sigma\Sigma$ takes place\cite{Jaqaman01,Gal00}.
We see that at high temperatures the $\Lambda$ is more abundantly produced with the set (Ia)
than when the set (II) is used. The $\Sigma$ has the opposite behaviour, as its
abundance obtained from the set (Ia) is less than that from the set (II).
At the same time, there is no noticeable difference between the two models 
in the nucleon, kaon and cascade production rates.
Furthermore, there is not much difference in total strange baryon production.
Generally speaking, it can be concluded that kaon production and condensation 
in the context of MQMC is not likely 
to be affected drastically by which set is being used from amongst the different sets of 
parameters extracted from phenomenology. 

\begin{figure}
\includegraphics{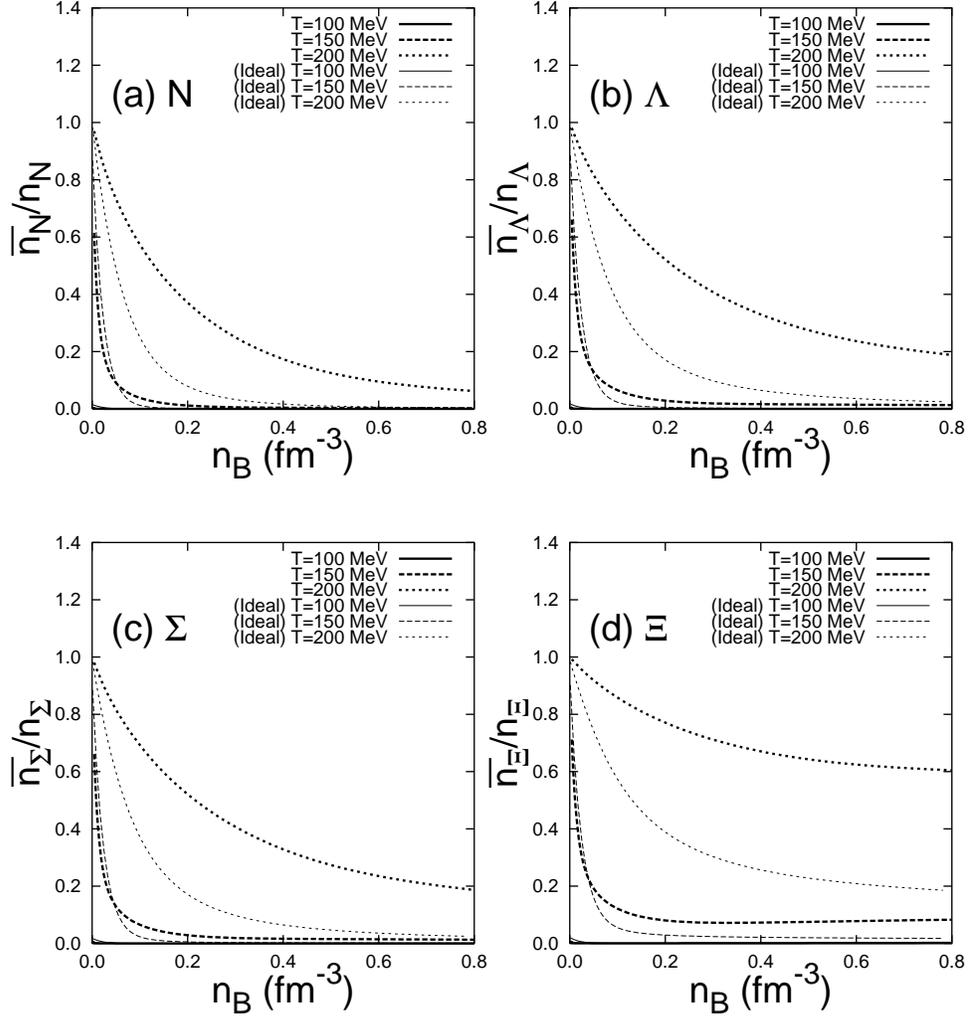}%
\caption{\label{fign3}
The ratio of anti-baryon to baryon abundance
$\overline{n}_i/n_i$ for each baryon species 
versus baryonic density $n_B$ at several temperatures using the parameter
set (Ia) and for the ideal gas approximation.
a) N, b) $\Lambda$, c) $\Sigma$, d) $\Xi$.}
\end{figure}

\begin{figure}
\includegraphics{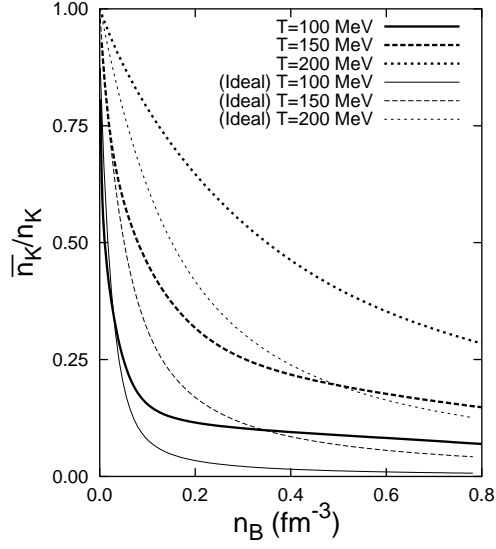}%
\caption{\label{fign4}
The ratio of anti-kaon to kaon abundance
$\overline{n}_K/n_K$ at various temperatures using the parameter set (Ia)
and for the ideal gas approximation.}
\end{figure}

In Fig.(\ref{fign3}), we display the anti-baryon to baryon abundance ratio 
for each baryon species $\overline{n}_i/n_i$ 
at several temperatures as obtained by using parameter set (Ia). 
It is seen that for temperatures $T < 100 \mbox{MeV}$ the production
of the anti-baryons is negligible. 
However, for higher temperatures   
there is significant anti-baryon production
particularly when the temperature reaches and exceeds $T_c$. 
Furthermore, we notice that the 
anti-baryon abundance ratio
$\overline{n}_i/n_i$, is particularly large at small baryon density $n_B$. 
This ratio also seems to increase with the baryon strangeness number as follows:  
\begin{eqnarray}
\frac{\overline{\Xi}}{\Xi}>
\frac{\overline{\Lambda}}{\Lambda}, \frac{\overline{\Sigma}}{\Sigma}> 
\frac{\overline{N}}{N}.
\end{eqnarray}
This enhancement of anti-hyperon abundance ratio over the antiproton abundance ratio 
in the relativistic nuclear collisions has been observed earlier\cite{Schaffner91}. 
Also included in Fig.(\ref{fign3}) are the results obtained in the ideal gas approximation
which assumes noninteracting hadrons in a thermal bath. In this approximation,
the hadrons do not modify their masses, do not interact with each other and
there are no mean fields at all. Although this approximation is too simplistic, 
it can be useful to obtain rough first estimates. It is seen from Fig.(\ref{fign3}) 
that the ideal gas results greatly 
underestimate the anti-baryon production, 
especially at high temperature or at high density.
In Fig.(\ref{fign4}), 
we show the  anti-kaon to kaon abundance ratio 
versus the baryon density $n_B$ at various temperatures using parameter 
set (Ia) and in the ideal gas approximation.
We note that the anti-kaon abundance ratio increases with temperature.
We also note that again the ideal gas approximation gives much lower estimates 
for the anti-kaon to kaon abundance ratio than the MQMC, especially at 
high temperature or at high density.

\begin{figure}
\includegraphics{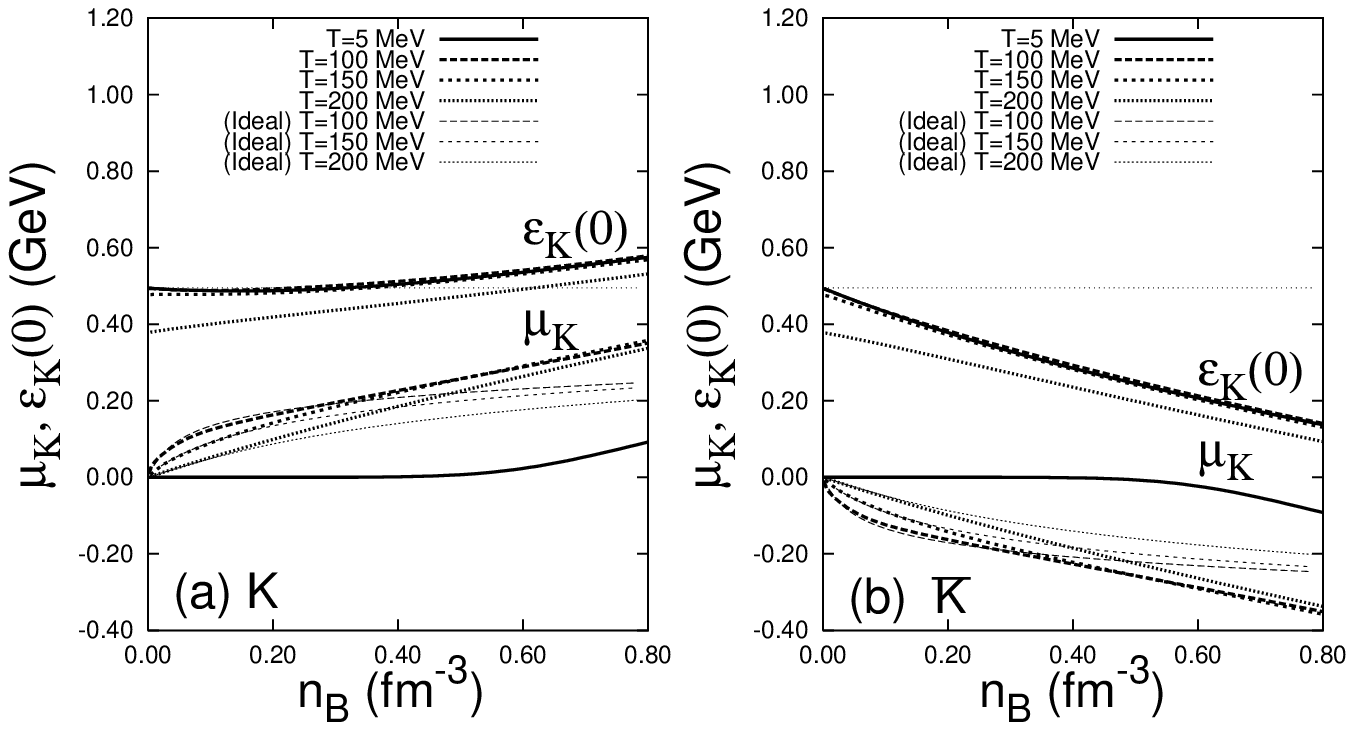}%
\caption{\label{fign5}
The strange chemical potential $\mu_S$ and 
the kaon threshold effective energy $\epsilon^{*}_k(0)$ 
versus baryonic density $n_B$ for various temperatures 
using the parameter set (Ia).
a) Kaons, b) Anti-kaons.}
\end{figure}

Fig.(\ref{fign5}) examines the crucial issue of the onset of kaon condensation in heavy ion collisions. 
Kaon condensation implies that the equation of state will be softer and that more hyperons
and kaons will accumulate in the system. Indeed, it is known that kaon condensation 
plays a significant role in the physics of neutron stars\cite{Thorsson94}. 
One of the major themes of the present work is to study the possibility 
for the onset of kaon condensation in heavy ion collisions.
The necessary condition for the onset of kaon condensation is to nontrivially satisfy 
Eq.(\ref{c_condensate}).  
For the $K^0$ or $K^+$ kaons, condensation is possible only when 
\begin{eqnarray}
\mu_k=\epsilon^{*}_k(0)
\label{c_kaon}
\end{eqnarray}
while for the the antikaons $\overline{K}^0$ or $K^-$, condensation is possible only when 
\begin{eqnarray}
\mu_k=-\overline{\epsilon^{*}_k}(0).
\label{c_antikaon}
\end{eqnarray}
When the condition for the onset of kaon condensation is satisfied, the kaons 
start to condense
smoothly and the strange chemical potential could be evaluated either from
Eq.(\ref{c_kaon}) or (\ref{c_antikaon}). 
After the kaon condensation starts to take place in the system, 
the strange chemical potential $\mu_S$ 
becomes an input parameter and the amplitude of the kaon condensation is calculated 
by conserving the total strangeness of the system. 
The contribution of the kaon condensate to the total density of the system 
is calculated from Eq.(\ref{c_amplitude}).
To examine the possibility for the onset of kaon condensation in our model 
we plot in Fig.(\ref{fign5}) the kaon chemical potential $\mu_K$ 
and the kaon threshold effective energy $\epsilon^{*}_k(0)$ 
versus the baryonic density $n_B$ at various temperatures.
The isospin chemical potential $\mu_Q$ vanishes for symmetric 
strange hadronic matter so that in this case Eq.(\ref{k_chemical}) reduces to 
 $\mu_K=\mu_S$ where $\mu_S$
is the strangeness  chemical potential. 
We see from panel (a) of Fig.(\ref{fign5}) that the kaon chemical potential $\mu_K$ 
is always less than the kaon threshold energy $\epsilon^{*}_k(0)$
for the all baryon densities $n_B$ so that Eq.(\ref{c_kaon}) can never be satisfied. 
A similar conclusion can be drawn for the 
anti-kaons and Eq.(\ref{c_antikaon}) as can be seen from panel (b) of the same figure. 
This indicates that kaon condensation does not take place at any temperature and density
in heavy ion collisions 
where the total strangeness is conserved to zero for a system that consists of
baryons and kaons. The kaons are only produced 
thermally in this system. 
The situation is, of course, completely different in 
neutron stars where there is enough time for the conservation of strangeness 
to be violated by the weak interaction\cite{Thorsson94}.

\begin{figure}
\includegraphics{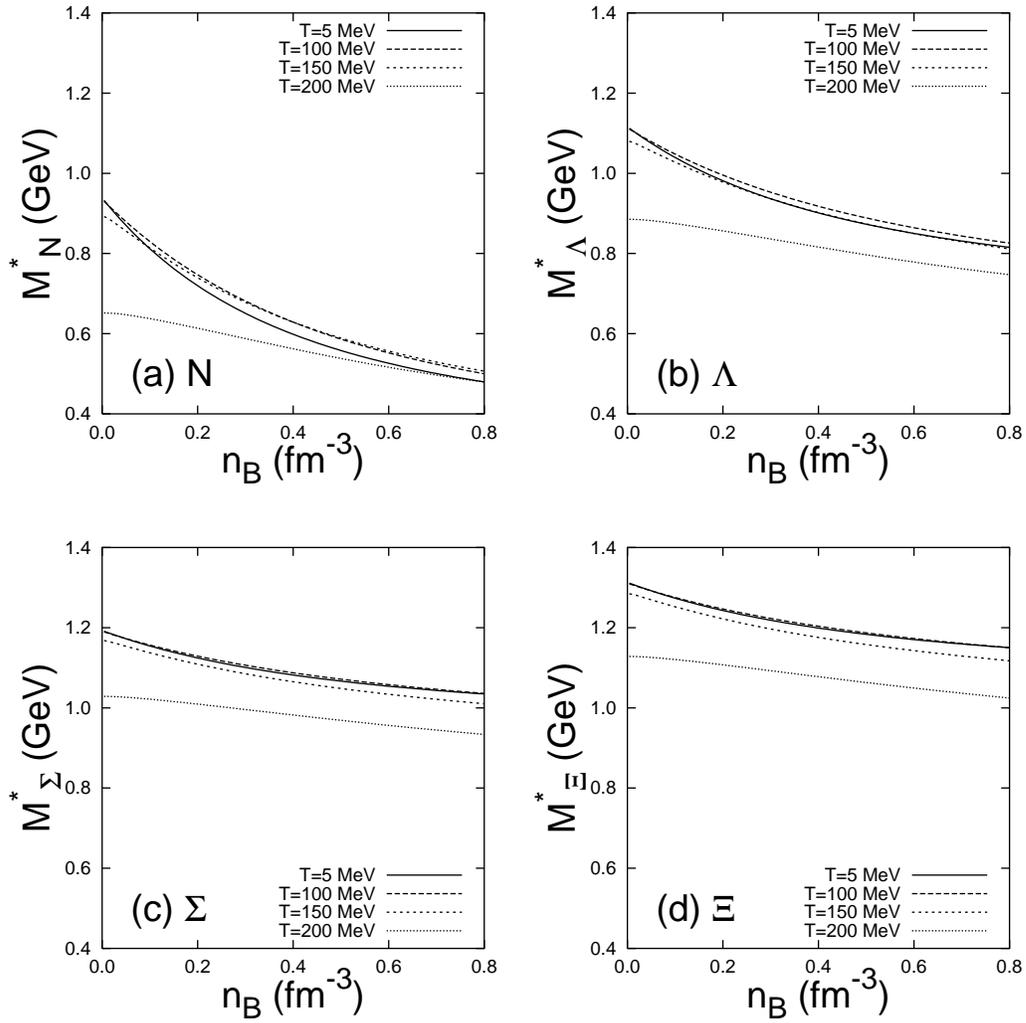}%
\caption{\label{fign6}
The effective mass $M^{*}_i$ for each baryon species versus baryonic 
density $n_B$ for various temperatures using the parameter set (Ia). 
a) $N$, b) $\Lambda$, c) $\Sigma$, d) $\Xi$.}
\end{figure}

\begin{figure}
\includegraphics{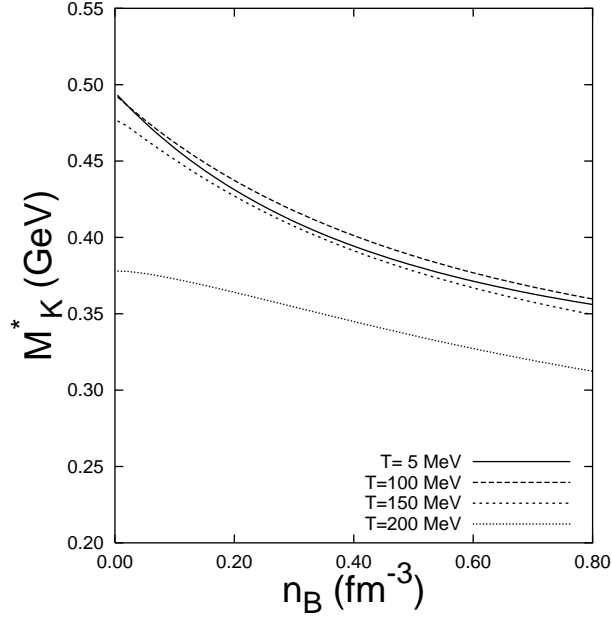}%
\caption{\label{fign7}
The effective kaon mass $M^{*}_K$  versus baryonic density $n_B$
for various temperatures using the parameter set (Ia).}
\end{figure}

In Figs.(\ref{fign6}) and (\ref{fign7}), we display the baryon and kaon 
effective masses versus the baryon density $n_B$. 
We note that, in the context of the MQMC model, the effective masses do 
not vanish at large baryon densities, unlike the standard Walecka model 
where the effective hadronic masses tend 
to vanish at large baryon density. The realistic case of effectively 
massive hadrons at all densities 
is actually a characteristic result of the MQMC.  
Indeed, in the hadronic phase, the hadron's effective mass should not vanish until chiral 
symmetry is restored or a QGP phase transition takes place. 
At the phase transition, there is a sharp drop in the hadron's effective
masses and all hadrons in the system will dissolve into their constituent quarks.
Furthermore, the quarks attain their current masses which are very small for 
the up and down flavors. 
At this point, however, bubbles of the QGP are formed in the system. 
Therefore, we do not find it is appropriate, in 
the realistic situation relevant to heavy ion collisions, 
to extrapolate the study of the onset of kaon condensation to massless hadrons since these
massless hadrons can exist only at the point of the phase transition.

\begin{figure}
\includegraphics{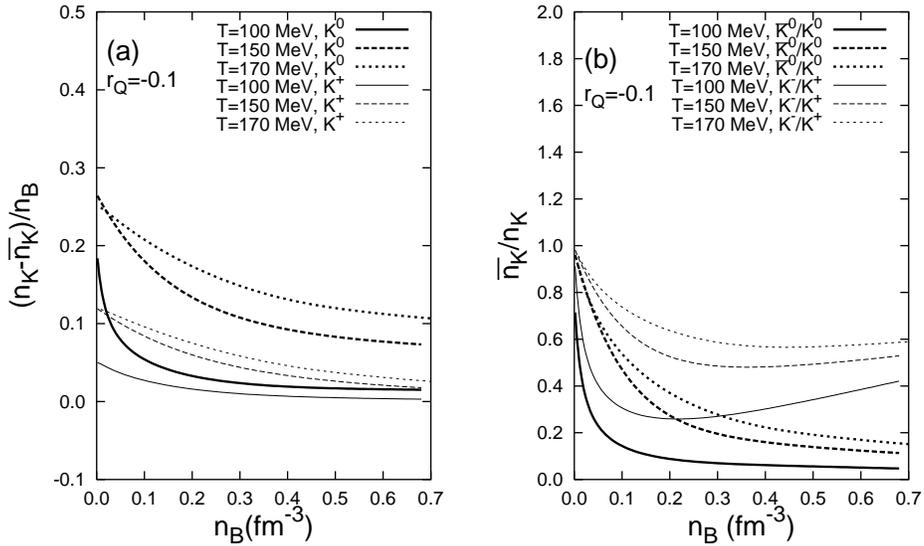}%
\caption{\label{fign8}
The kaon abundance versus baryonic density $n_B$ for various
temperatures for $n_Q/n_B=-0.1$ using the parameter set (Ia).
a) The kaon density fraction $x_K=(n_{K}-\overline{n}_K)/n_B$,
b) The ratio of anti-kaon abundance over kaon abundance
$\overline{n}_K/n_K$.}
\end{figure}

\begin{figure}
\includegraphics{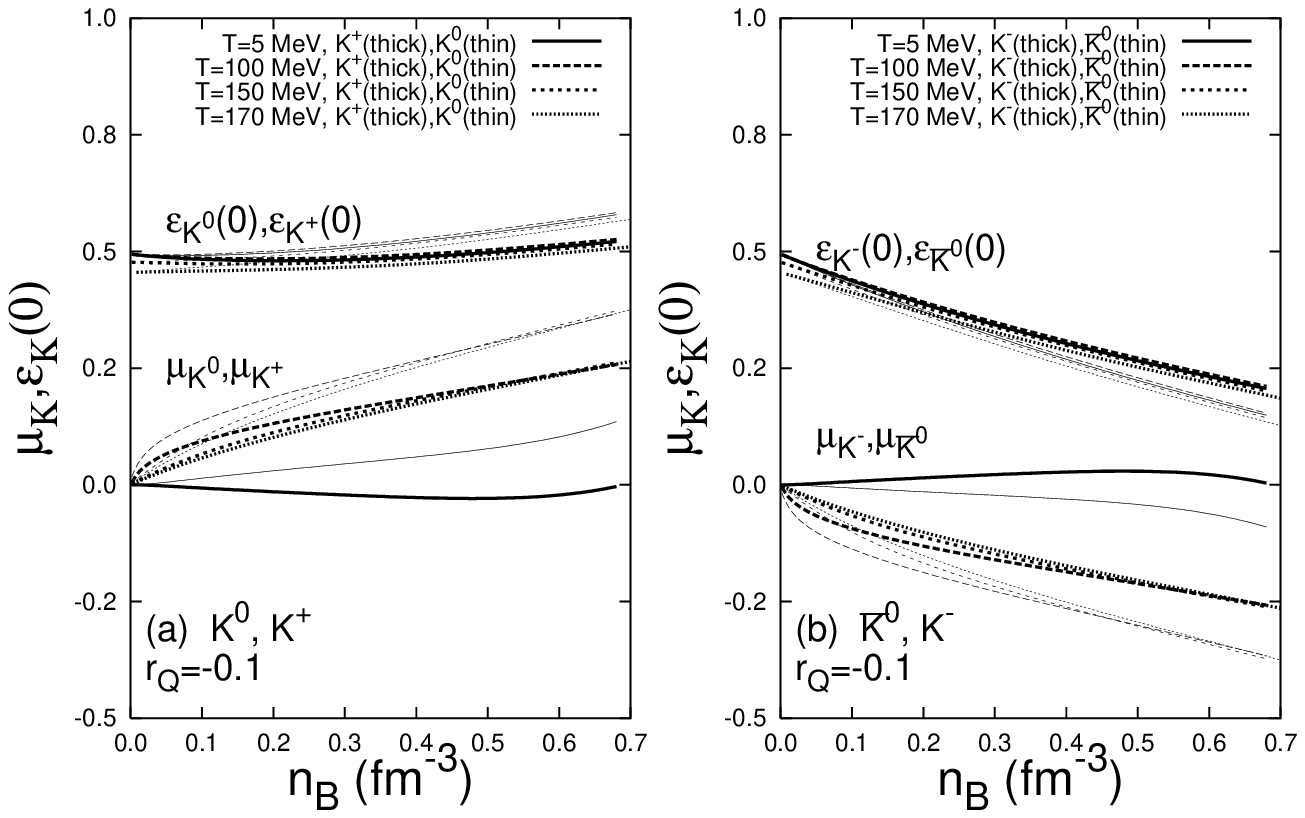}%
\caption{\label{fign9}
The strange chemical potential $\mu_S$ and
the kaon threshold effective energy $\epsilon^{*}_k(0)$
versus baryonic density $n_B$ at various temperatures
for $n_Q/n_B=-0.1$ using the parameter set (Ia).
a) Kaons, b) Anti-kaons.}
\end{figure}

\begin{figure}
\includegraphics{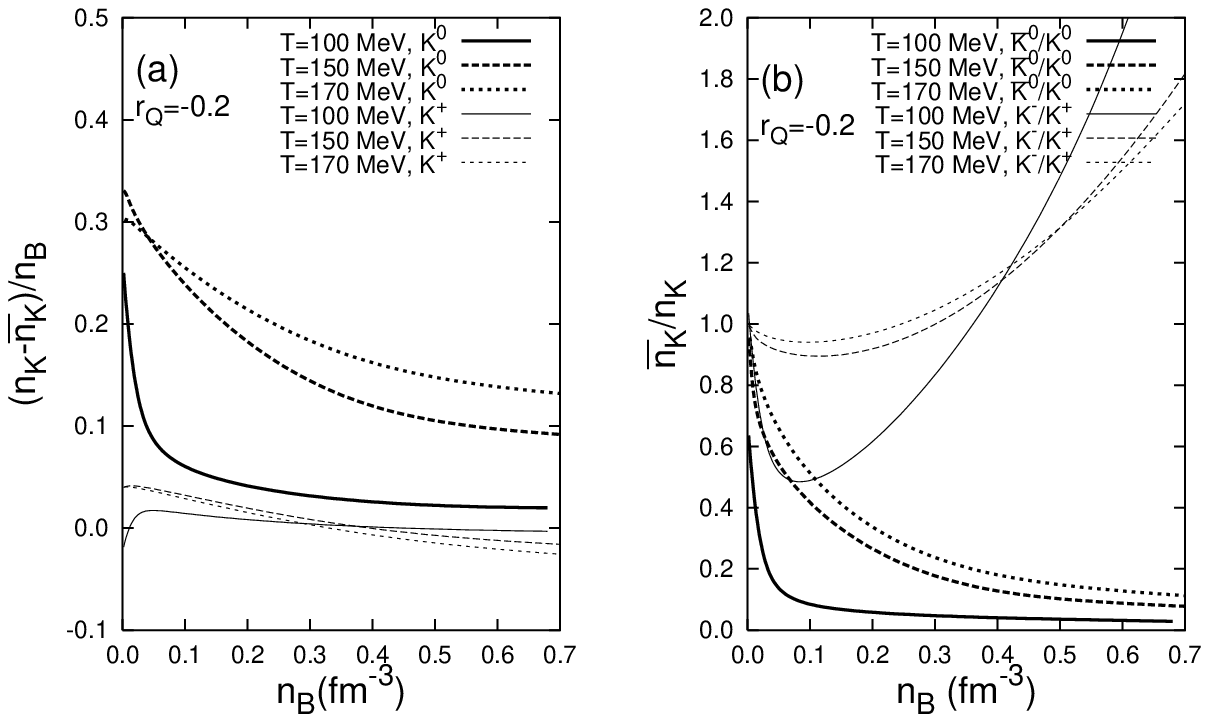}%
\caption{\label{fign10}
Same as Fig.(\ref{fign8}) but with $n_Q/n_B=-0.2$.}
\end{figure}

\begin{figure}
\includegraphics{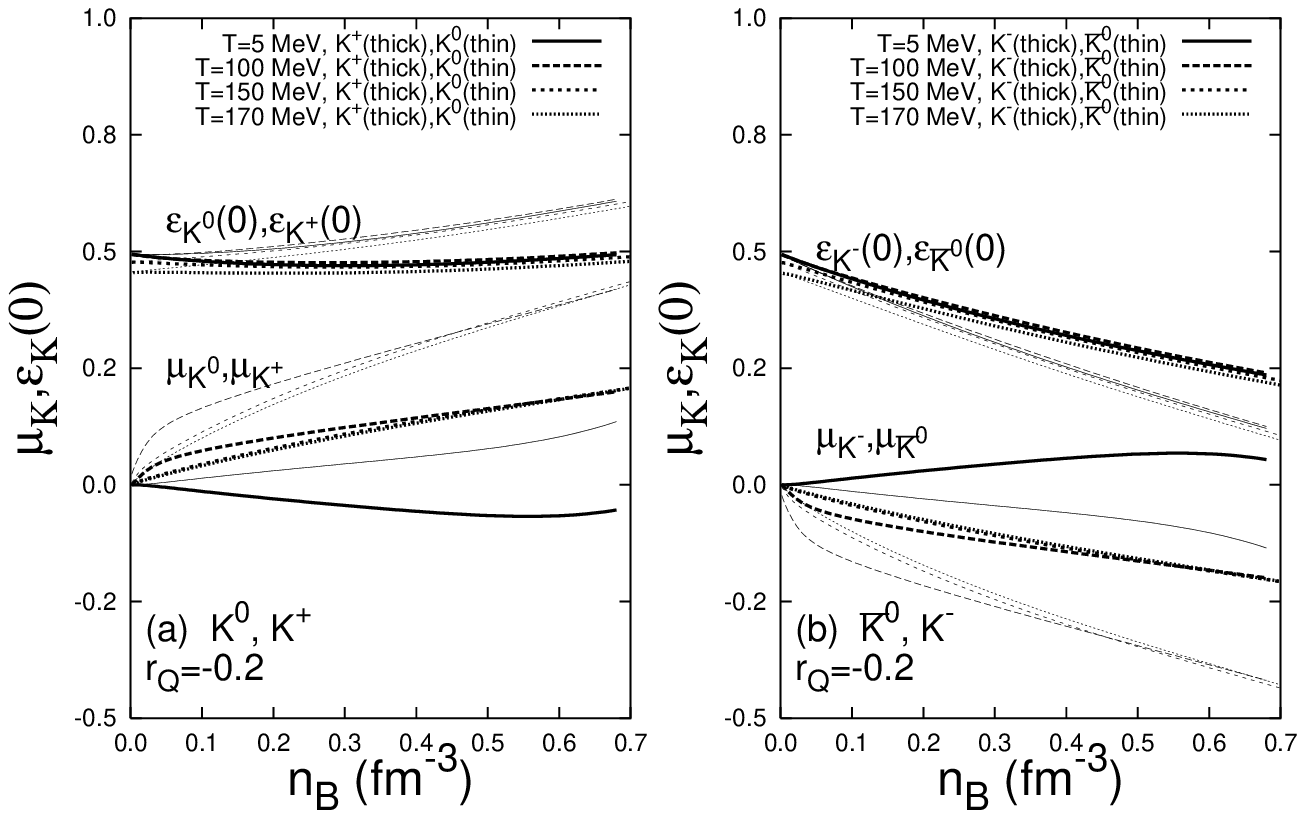}%
\caption{\label{fign11}
Same as Fig.(\ref{fign9}) but with $n_Q/n_B=-0.2$.}
\end{figure}

We have also extended MQMC to study hot and dense asymmetric strange hadronic matter 
with a conserved total zero strangeness $n_S/n_B=0$ and a conserved finite 
negative isospin density fraction $n_Q/n_B\le 0$.
Generally speaking, the negative value of the isospin density fraction $n_Q/n_B$ 
reflects the fact that the number of neutrons is more than the number of protons
(i.e. $n_n > n_p$) in the initial conditions of the colliding heavy ions.
The results for the asymmetric system are therefore more relevant to 
relativistic heavy ion collisions. We present our results for such a system 
using the parameter set (Ia) in Figs. (\ref{fign8}) and (\ref{fign9})
for $n_Q/n_B = -0.10$, while the results for $n_Q/n_B = -0.20$ 
are shown in Figs.(\ref{fign10}) and (\ref{fign11}).

In Figs.(\ref{fign8}) and (\ref{fign10}),
we display the net kaon abundance ratio $x_K=(n_K-\overline{n}_K)/n_B$
and the ratio of anti-kaon abundance over kaon abundance
for the charged and neutral kaon species $(K^+,K^0)$ 
at several temperatures for the isospin charge density fractions 
$n_Q/n_B=-0.1$ and $n_Q/n_B=-0.2$, respectively.
In panel (a) of both figures, we display the net kaon abundance 
versus the baryon density $n_B$.
It is seen that $x_{K^0}=(n_{K^0}-n_{\overline{K}^0})/n_B$ is much larger than
$x_{K^+}=(n_{K^+}-n_{K^-})/n_B$ in such a system, especially at higher temperatures.
The conservation of net strangeness for the asymmetric system is more easily
preserved by neutral kaons at high temperatures.
Consequently, when the negative isospin $n_Q/n_B$ is increased,
the neutral kaons become more dominant than the charged ones.
>From Fig.(\ref{fign8}.b), we see that
the $K^0$ and $K^+$ are produced more abundantly than their 
anti-particles for the system with lower asymmetry.
However, when the asymmetry is increased, we see from Fig.(\ref{fign10}.b) 
a dramatic change for the charged kaons with the anti-particle $K^-$ becoming more 
abundant than the  $K^+$ at high densities in order to conserve isospin. 
Actually, in addition to the kaons, 
we have found that the baryon species $\Sigma^-,\Sigma^0,\Sigma^+$ 
also play an important role in conserving the isospin $n_Q/n_B$ 
at high temperature when the isospin charge fraction is increased.
The proton density fraction tends to saturate 
at some value when the negative isospin charge fraction
becomes rather large ($n_Q/n_B=-0.2, n_p/n_B\approx 0.30$)
while the density fraction for neutrons continues to increase with temperature.
However, at higher temperatures, the conservation of isospin charge fraction 
is preserved by creating more charged strange hadrons. 

We have also examined the possibility for the onset of kaon 
condensation in the asymmetric system.
The necessary condition for kaon condensation is that the kaon threshold energy
$\epsilon^{*}_k(0)$ equal its chemical potential $\mu_K$.
In Figs(\ref{fign9}) and (\ref{fign11}), we display the kaon chemical potential $\mu_K$ 
and the kaon threshold effective energy $\epsilon^{*}_k(0)$
versus the baryonic density $n_B$ at various temperatures and
for $n_Q/n_B=-0.1$ and $n_Q/n_B=-0.2$, respectively.
We see that the kaon's threshold energy is always larger than its chemical potential, and 
their curves become almost parallel at large baryonic density.
Furthermore, when the nuclear matter asymmetry is increased by increasing 
the negative isospin $n_Q/n_B$
to reach the maximal value of $n_Q/n_B=-0.5$, appropriate for neutron matter,
the onset of kaon condensation is not possible even at very large baryonic densities 
and we find that the strange chemical potential $\mu_S$ in the medium is always modified 
in such a manner as to avoid the onset of kaon condensation.
These results show that no signature can be expected for kaon condensation 
in heavy ion collisions and kaons can only be produced thermally. 

\begin{figure}
\includegraphics{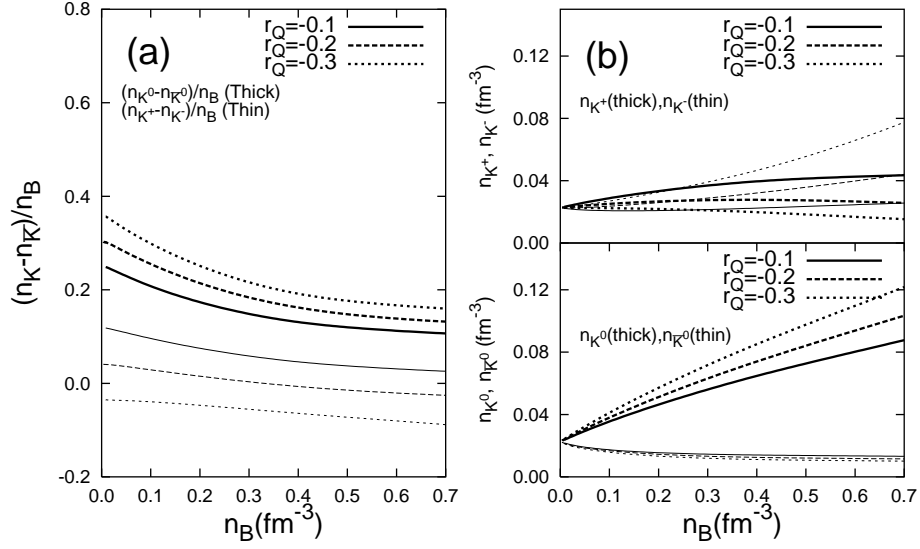}%
\caption{\label{fign12}
The kaon and anti-kaon abundances versus baryonic density $n_B$
for various isospin charge ratios $n_Q/n_B$ at temperature $T=170$ MeV using the parameter set (Ia).
a) The density fraction $x_K=(n_{K}-n_{\overline{K}})/n_B$ for the charged and neutral kaons,
b) The kaon and anti-kaon abundances for charged (upper panel) and neutral (lower panel) kaons.} 
\end{figure}

We have also examined how the production of the charged and neutral kaons 
and their anti-particles at high temperature varies with respect 
to the isospin charge ratio $n_Q/n_B$.
In fig.(\ref{fign12}), we display the kaon and antikaon abundances versus 
the baryonic density $n_B$ at temperature $T=170$ MeV 
for $n_Q/n_B=$ -0.1, -0.2, -0.3.
The temperature $T=$ 170 MeV is particularly interesting 
because it is accessible at RHIC energies and is relevant 
to the freezeout chemical potential \cite{Braun-Munzinge01,Zschiesche02}.
Part (a) of fig.(\ref{fign12}) displays the charged and neutral kaon abundant density fractions, 
$x_K$, while part (b) displays the densities for charged (upper panel) 
and neutral (lower panel) kaons and their antiparticles. 
It is seen from part (b) that the kaons and their antiparticles 
are produced abundantly at high temperatures even at very low baryon densities.
However, the net density fraction for charged kaons $x_K=(n_{K^+}-n_{K^-})/n_B$ 
tends to be small and varies weakly with the baryon density $n_B$.
Furthermore, it decreases when the asymmetry of the system increases
by increasing the negative value of the isospin charge fraction $n_Q/n_B$.
This reflects the fact that the negative charged kaons become 
more dominant as the magnitude of the isospin charge asymmetry increases 
and exceeds $|n_Q/n_B|=0.25$.
Here it must be stressed that in a realistic situation the isospin charge fraction 
of the normally used colliding heavy ions lies in 
the range $n_Q/n_B=0$ to $-0.1$ and consequently 
$K^+$ production is expected to be more dominant than $K^-$ in RHIC.
However, more asymmetric nuclear matter may be reached 
experimentally in the future by using different isotopes 
for the colliding heavy ions.
Generally speaking, the net production of charged kaons is relatively 
smaller than that of the neutral kaons, indicating that 
the neutral kaons play a more significant role
in  asymmetric nuclear matter in order to conserve the net zero strangeness 
of the system.

\begin{figure}
\includegraphics{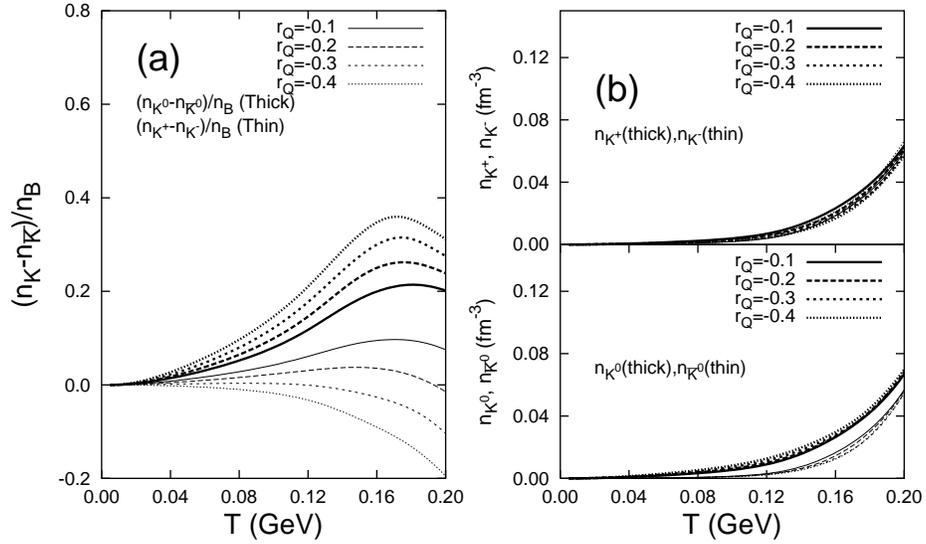}%
\caption{\label{fign13}
Same as Fig.(\ref{fign12}) but versus temperature at the baryonic density $n_B=0.05 \mbox{fm}^{-3}$.}
\end{figure}

Fig.(\ref{fign13}) displays the variation 
of kaon and antikaon abundances with respect to the temperature 
for several values of the isospin charge fraction $n_Q/n_B$. 
The calculations are carried out 
at the baryonic density $\rho_B = 0.05 \mbox{fm}^{-3}$.
This rather small baryonic density is accessible in RHIC 
and has a special interest at high temperature in particular in  
studying the freezeout chemical potential \cite{Braun-Munzinge01,Zschiesche02}.
The phase transition from the hadronic phase to QGP
is expected to take place at this low baryon density 
and temperature $T=170 \mbox{MeV}$.
It is seen that kaons are produced 
abundantly when the temperature exceeds $T>120 \mbox{MeV}$ 
and this production becomes more significant 
when the temperature reaches $T=170 \mbox{MeV}$.
Furthermore, it is seen that the neutral kaon density, $n_{K^0}$, 
lies always above the antikaon density $n_{\overline{K}^0}$
but this is not the case for the charged kaons, $n_{K^+}$ and $n_{K^-}$. 
The net density fraction for the charged kaons decreases 
as the magnitude of the negative isospin charge asymmetry increases
while net density fraction for the neutral kaons has the opposite behavior.
Moreover, the net density for the neutral kaons is always greater 
than that for the charged kaons for asymmetric nuclear matter.
This reflects the fact that the $K^0$ abundance is dominant at all temperatures
and more so as the magnitude of the isospin charge asymmetry increases.

\begin{figure}
\includegraphics{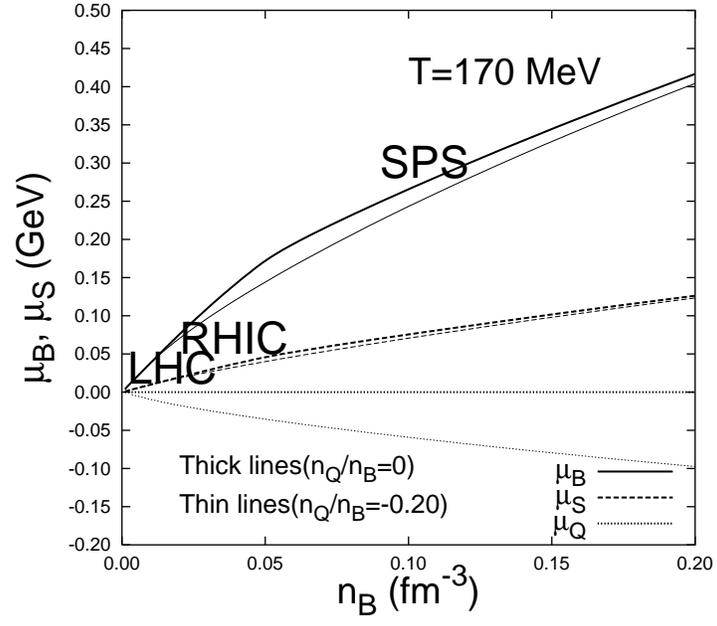}%
\caption{\label{fign14}
The dependence of baryon, strange and isospin chemical potentials $\mu_B$, $\mu_S$ and $\mu_Q$, respectively, 
on the baryon density $n_B$ at $T=170 \mbox{MeV}$ for a symmetric and an asymmetric matter.}
\end{figure}

In Fig.(\ref{fign14}), we display the baryon, strangeness and isospin 
chemical potentials versus the baryonic density $n_B$ 
at $T=170 \mbox{MeV}$ for the symmetrical and asymmetrical nuclear matter. 
We also indicate on this figure the estimated
freezeout chemical potential at SPS and RHIC energies.
In particular it is known that freezeout in RHIC occurs at the small chemical 
potential $\mu_B\approx 46\pm 5 \mbox{MeV}$ 
and at a temperature $T\approx 174\pm 7 \mbox{MeV}$ 
\cite{Braun-Munzinge01,Zschiesche02}. 
The substantial decrease of the baryon chemical potential 
from $\mu_B\approx 270 \mbox{MeV}$ at SPS energies to
$\mu_B\approx 45 \mbox{MeV}$ at RHIC energies shows that, at mid-rapidity, 
we are dealing with a low net baryon density
in the medium\cite{Braun-Munzinge01}. In our model, the baryon chemical 
potential $\mu_B\approx 50 \mbox{MeV}$ 
for the critical temperature $T\approx 170 \mbox{MeV}$ corresponds 
to the very small baryonic density 
$n_B\approx 0.015 \mbox{fm}^{-3}$. 
Other models, however, may lead to somewhat higher values for the baryonic 
density for the chemical freezeout potential\cite{Zschiesche02} and
we shall refer to a baryon density  $n_B\approx 0.05 \mbox{fm}^{-3}$ 
when we speak about RHIC physics. 

%

%

\section{Conclusions and Summary}

We have studied strange hadronic matter composed of the baryon octet and kaon doublet 
with a conserved zero total strangeness 
$n_S/n_B=\left(n^{\mbox{Baryons}}_S+n^{\mbox{Kaons}}_S\right)/n_B=0$ 
and a conserved finite isospin fraction $n_Q/n_B$.
RHIC physics is reached at the rather low baryonic density $n_B\approx 0.05 \mbox{fm}^{-3}$
and high temperature $T \approx 150-170 \mbox{ MeV}$\cite{Braun-Munzinge01}.
We have not found any evidence for kaon condensation 
even in highly asymmetric dense nuclear matter. 
The kaon threshold energy is always higher than the kaon chemical potential 
given by Eq.(\ref{k_chemical}) so that the kaons are only produced thermally in the system.
The production of the strange hadrons starts smoothly at low temperatures 
and then they are abundantly produced when the temperature reaches $T\approx 170 \mbox{MeV}$.
 
We have solved the baryon, strange and isospin chemical potentials 
$\mu_B, \mu_S$ and $\mu_Q$, respectively, self-consistently. 
In the hot or even warm dense medium, the strange chemical potential $\mu_S$ 
is finite (nonzero) even for a system with zero total  strangeness and 
it likely modifies itself in a manner to decrease the kaon chemical 
potential to be always lower than the kaon threshold energy.
The results of other microscopic models also disfavour kaon with respect to 
pion condensation\cite{Barducci04}. 
On the other hand, in cold dense matter, the strange chemical potential 
$\mu_S$ is normally set to zero for a system with zero total  strangeness. 
In this case, it is possible for the antikaon (kaon) to condense at zero temperature 
at very high baryonic density. 
Therefore, the results at finite temperature suggest that $\mu_S$ must be actually assigned  
a small finite value at zero temperature
in order to maintain the conservation of the zero total strangeness for cold dense matter.
Such a modification would also reduce the possibility for 
the onset of kaon condensation at $T=0$.

It is interesting to note here that the effective hadronic masses 
do not vanish in the hadronic phase in our model. 
The effective hadronic masses attain small values only at the chiral phase transition 
and the phase transition to the QGP.
In the QGP,  the hadrons dissolve and disappear and the kaons will not be produced. 
On the other hand, at very high densities and low temperatures the quark color 
superconductivity will dominate.
The strange hadrons and their anti-particles are normally produced at high 
temperatures and low baryonic densities
which are the ideal circumstances at  RHIC energies. 
Furthermore, when the system freezes out, the strange baryons and kaons might survive with
the baryonic sector having a negative net strangeness 
$n^{\mbox{baryons}}_S/n_B=0.0$ to $-0.25$ 
and the mesonic sector (i.e. kaons)having a positive net strangeness,
but the total strangeness for the baryons and kaons is conserved to zero. 
Moreover, although $K^0$ production always  dominates over $\overline{K}^0$,
the $K^{-}$ becomes more abundant than $K^{+}$ 
for highly asymmetric nuclear matter (i.e.: $n_Q/n_B<-0.2$), especially at high temperatures or 
large baryonic densities.

\begin{acknowledgments}
I. Z. gratefully acknowledges support from the Alexander von Humboldt Foundation.
The authors are indebted to the Deutsche Forschungsgemeinschaft 
for financial support during
the early stages of the present project through the grant GR 243/51-1.
I. Z. and H. R. J. thank G. Shahin for his assistance in the computational work. 
I.Z. thanks C. Greiner, J. Schaffner-Bielich, L. Tolos and J-P. Leroy for interesting  
discussions and comments.
\end{acknowledgments}

\bibliography{manheim_henry1.bib}

\newpage
\newpage
\end{document}